\documentclass[11pt,letterpaper,notoc,nohyper]{JHEP3}
\pdfoutput=1
\usepackage{graphics}
\usepackage{amsmath}
\usepackage{cite}
\usepackage{colordvi}
\usepackage{epsfig}
\usepackage{latexsym}

\newcommand{\slashchar}[1]{\setbox0=\hbox{$#1$}   
   \dimen0=\wd0                                     
   \setbox1=\hbox{/} \dimen1=\wd1                   
   \ifdim\dimen0>\dimen1                            
      \rlap{\hbox to \dimen0{\hfil/\hfil}}          
      #1                                            
   \else                                            
      \rlap{\hbox to \dimen1{\hfil$#1$\hfil}}       
      /                                             
   \fi}                                             %

\newcommand{\ETmiss}{\slashchar{E}_T}
\newcommand\nn{\nonumber}
\newcommand\gev{\text{GeV}}
\newcommand\tev{\text{TeV}}
\newcommand\pb{\text{pb}}
\newcommand\fb{\text{fb}}
\newcommand\jet{\text{jet}}
\newcommand\jets{\text{jets}}

\title{\mbox{Mass-Matching in Higgsless}}
\author{
Adam Martin$^a$\thanks{adam.martin@yale.edu}  and Veronica Sanz$^{b,c}$\thanks{vsanz@bu.edu} \\ 
\llap{$^a$} Department of Physics, Sloane Laboratory, Yale University, 
New Haven, CT 06520 \\
\llap{$^b$} Department of Physics, Boston University, 
Boston, MA 02215 \\
\llap{$^c$} Department of Physics and Astronomy, York University, Toronto, Ontario, Canada 
}

\abstract{Modern extra-dimensional Higgsless scenarios rely on a mass-matching between fermionic and bosonic KK resonances to evade constraints from precision electroweak measurements. After analyzing all of the Tevatron and LEP bounds on these so-called Cured Higgsless scenarios,  we study their LHC signatures and explore how to identify the mass-matching mechanism, the key to their viability. We find singly and pair produced fermionic resonances show up as clean signals with 2 or 4 leptons and 2 hard jets, while neutral and charged bosonic resonances are visible in the dilepton and leptonic $W^{\pm}Z^0$ channels, respectively. A measurement of the resonance masses from these channels shows the matching necessary to achieve $S\simeq 0$. Moreover, a large single production of KK-fermion resonances is a clear indication of compositeness of SM quarks. Discovery reach is below 10 fb$^{-1}$ of luminosity for resonances in the 700 GeV range. }
\begin{document}



\section{Introduction}
 
 Tantalizing though it is, the idea that electroweak symmetry could be broken without a Higgs particle has proven hard to reconcile with data. Such dynamical symmetry breaking assumes that some non-perturbative effects produce composite scalar states which mix with the $W^{\pm}$ and $Z^0$ to give them masses. In  technicolor~\cite{Susskind:1978ms, Weinberg:1979bn}, the non-perturbative effects arise from strong forces analogous to those mediated by the gluons of QCD -- in which a phase transition causes quarks to be confined within pions.

Assuming non-perturbative dynamics brings a major drawback. Extraction of  measured quantities from a top-down approach is often unmanageable. Particularly difficult is the $S$-parameter. Estimates of $S$ in technicolor models based on rescaling QCD predict large and positive values of $S$~\cite{Peskin:1990zt, Golden:1990ig,  Holdom:1990tc}, whereas experiments favor $S \simeq 0$. The $S$ parameter is not directly measured in experiments, but correlated with other pseudo-observables such as the $T$ parameter. Pushing toward non-zero values of $T$ does afford wiggle in the allowed value of $S$, but not nearly enough to accommodate QCD-like values of $S$\@. There are several arguments that the $S$ parameter should naturally be lower in technicolor models which exhibit near-conformal ('walking'~\cite{Holdom:1981rm, Yamawaki:1986zg, Appelquist:1986an}) behavior~\cite{hep-ph/9206225, Lane:1994pg, Appelquist:1998xf, Kurachi:2006mu}. However as there is no systematic calculation of $S, T$ in non QCD-like technicolor, precision electroweak compatibility  is still the biggest hurdle facing technicolor.

Naturally achieving $T \sim 0$ is on a different footing from $S \sim 0$: it can be enforced by implementing custodial symmetry into any model\cite{Sikivie:1980hm,Terning:1994sc,Agashe:2003zs}. On the other hand, there is no known symmetry protecting $S$\@.
In the absence of a symmetry, traditional non-perturbative approaches in technicolor have had little to say about the possibility of $S \sim 0$. More recently however, holography has provided insight into suppression mechanisms which are not described in terms of symmetries. In holography, the 4D dynamics is mapped into a five-dimensional (5D) model. Localization of different fields in an extra dimension is interpreted in 4D as dynamical effects from the non-perturbative dynamics, where the renormalization group evolution is encoded into the wave-functions along the extra-dimension. 

While 5D models leave open the question of the details leading to these dynamical effects, they give us a bonus over traditional 4D scenarios. If it is possible to find a localization of fields in the extra dimension which suppresses some undesirable operators, one can correlate the localization assumption with some other effects, such as the spectrum and decay rates. While this technique does not offer insight on the mechanism or dynamics underlying a suppressed operator, it does allow us to predict the observable consequences. Examples of this idea at work abound: approaches to describe solutions to the gauge\cite{Randall:1999ee} and mass hierarchies\cite{Grossman:1999ra,Gherghetta:2000kr} and flavor problems\cite{Cacciapaglia:2007fw}, have all been addressed in the holographic picture as a consequence of localization inside the bulk, and not as a consequence of symmetries. 

The same holographic approach can be used to achieve $S \sim 0$, and this suppression has important bearing on the spectrum of a theory. Indeed, possible ways to obtain a small $S$ in 5D models involve either a direct modification of the spectrum of spin-1 fields (Holographic Technicolor~\cite{Hirn:2006nt, Hirn:2006wg}, or HTC) or a balance of spin-1 versus spin-1/2 particles (Cured Higgsless~\cite{Cacciapaglia:2004rb}, or CHL)\@.  Both proposals predict that the mechanism responsible of $S \sim 0$ implies a {\it mass-matching} between Kaluza-Klein (KK) towers of different particles~\footnote{4D Technicolor models have borrowed the the insight of the holographic approach to postulate mass-matching towers of axial and vector resonances~\cite{SekharChivukula:2008gz, Eichten:2007sx, Foadi:2007ue} in the context of `walking' technicolor\cite{Dietrich:2008up,Nunez:2008wi}.}. 

\begin{figure}[h!]
 \centering
 	\includegraphics[width=7.5cm, height=6.0cm]{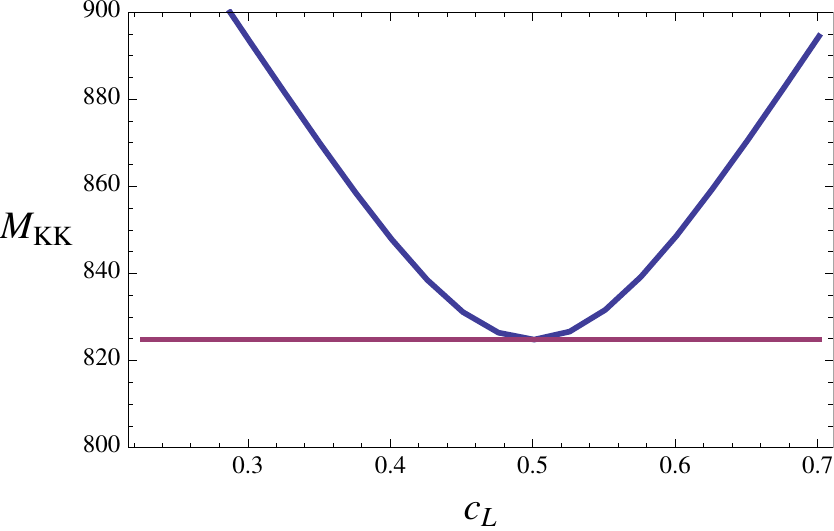}
\caption{Mass matching at work: the mass of the first $Z_{KK}$ state (red) and the mass of the first $Q_{KK}$ as a function of the 5D (left-handed) fermion mass $c_L$. Only values close to $c_L\sim 0.5$ agree with the current value of $S$.}
\label{fig:mm}
 \end{figure}

As a concrete example, in Figure~(\ref{fig:mm}) we show mass-matching at work in the Cured Higgsless scenario. Specifically, we show the mass separation between the first spin-1 resonance and the first fermionic resonance as a function of the left-handed fermions 5D mass, $c_L$. To achieve $S \sim 0$, this mass parameter must be chosen very close to particular value $c_L \sim 1/2$~\cite{Cacciapaglia:2004rb}, exactly where the splitting between the fermionic and gauge resonances is smallest. Thus, requiring $S\sim 0$ results in a mass-matching between the mesons (like the $Z_{KK}$, or techni-rho) and the baryons ($Q_{KK}$ or  techni-baryons) of the new strong interactions. The goal of this paper is to carefully examine the mass-matching mechanism. We study the implications of mass-mixing on the spectrum and explore how mass-matching may be deciphered from early LHC data.

The setup of this paper is the following: first, we briefly review the basics of 5D Higgsless models in Section (\ref{sec:hlessbasic}). The mass-mixing mechanism, which causes contributions to $S$ to cancel, is explained is Section~(\ref{sec:massmatch}). In Section~(\ref{sec:hlessPEW}) we explore the precision electroweak (PEW) and Tevatron constraints on mass-matched  Higgsless scenarios. Several collider signatures of the mass-matched models are then investigated in Section~(\ref{sec:hlessPHENO}). The three signatures we focus on are i.) Single production of fermion resonances, which is enhanced by the mass-matching mechanism,  ii.) Neutral KK resonances, specifically the visibility of multiple tiers of $Z_{KK}$ and its mass matching with the fermion resonance, and iii.) $W_{KK}$ resonances. Finally, in Section (\ref{sec:conclude}), we conclude. 

\section{Higgsless model basics}
\label{sec:hlessbasic}

In this section we review the basics of Higgsless models~\footnote{From this point on we will use `5D Higgsless' and `Higgsless' interchangeably. Purely 4D models of EWSB without a Higgs will be referred to as `deconstructed'.}.  The inspiration for Higgless models comes from the AdS/CFT correspondence~\cite{Gubser:1998bc,Witten:1998qj}; the observation that some strongly interacting systems can be qualitatively described by means of weakly coupled systems in higher dimensions\cite{Pomarol:2000hp,ArkaniHamed:2000ds}.  Higgsless  models are the implementation of this holographic idea for electroweak symmetry breaking (EWSB)\cite{Csaki:2003dt,Csaki:2003zu,Nomura:2003du}. Similar applications of this duality have been developed for various physical systems, including low-energy QCD\cite{Hirn:2005nr,Hirn:2005vk}, heavy-ion physics\cite{Policastro:2002se}, superconductors\cite{Hartnoll:2008vx}, superfluidity\cite{Herzog:2008he}, fractional quantum hall effect\cite{Fujita:2009kw}, among others. 

Strong EW-scale dynamics is a vast metaclass of models, and presumably only a subset have a holographic descriptions. Although many holographic  models are based in AdS space, e.g. Higgsless, other proposals,  such as Holographic Technicolor, use non-AdS geometries.

The starting point of a Higgsless theory is a slice of AdS space; a finite extra-dimension with coordinate $z$ between two branes $z \in (\ell_0, \ell_1)$ with geometry $ds^2 = \frac{\ell_0^2}{z^2}(\eta_{\mu\nu} dx^{\mu}dx^{\nu} - dz^2)$. The brane at $\ell_0$ corresponds to the cutoff of the theory, while the brane at $\ell_1$ corresponds to the scale where conformal (and, as it will turn out, EW symmetry) is broken.  To model the chiral symmetry of the system, we include $SU(2)_L \otimes SU(2)_R \otimes U(1)_{X}$ gauge fields in the bulk of the extra dimension. For simplicity we set the $5D$ gauge couplings of the two $SU(2)$ symmetries to be equal to $g_5$, while the $U(1)_X$ gauge coupling is $\tilde g_5$.

In addition to the field content, we must specify the boundary conditions for the gauge fields at  $z = \ell_0$ and $z = \ell_1$. With that information, one can solve the system by KK decomposition. The result is a tower of 4D fields with identical quantum numbers but different masses. The lightest mode is interpreted as the SM particle (in this case, a gauge boson), while the heavier KK copies are interpreted as resonances. To provide a successful model of EWSB,  the ratio between the $W^{\pm}$ and $Z^0$ masses must be correct, and a $U(1)_{em}$ symmetry must remain unbroken at low energy to preserve a massless photon. These conditions can be met entirely by choosing appropiate boundary conditions~\cite{Csaki:2003dt, Csaki:2003zu}:
\begin{eqnarray}
\text{at}\ z &=& \ell_1: \partial_5(L^a_{\mu} + R^a_{\mu}) = 0,\quad L^a_{\mu} - R^a_{\mu} = 0,\quad \partial_5 B_{\mu} = 0 \nn \\
\vspace{0.2cm}
\text{at}\ z &=& \ell_0: \partial_5 L^a_{\mu} = 0,\quad R^{1,2}_{\mu} = 0,\quad \partial_5(g_5 B_{\mu} + \tilde g_5 R^3_{\mu}) = 0,\quad \tilde g_5 B_{\mu} - g_5 R^3_{\mu} = 0 
\end{eqnarray}

For a given $\ell_0$, the remaining parameters $\ell_1, g_5,\text{and}\ \tilde g_5$ are set by demanding three constraints: the lightest charged state must have mass $m_W$, the lightest (other than the photon) neutral state must have mass $m_Z$, and the coupling of the (flat) photon to charged fields be $e = \sqrt{4\pi\alpha_{em}}$. Imposing these conditions, the only remaining parameter in the gauge boson profiles is the overall normalization. This remaining coefficient is set by matching the 5D fermionic zero-mode couplings to the $W^{\pm}$ and $Z^0$ to the 4D charged current and neutral current couplings: 
\begin{equation}
\label{eq:match}
g_{SM} = g_5 \int_{\ell_0}^{\ell_1} \frac{\ell_0^4}{z^4}\phi^2_{f_L}(z)\phi_W(z) dz, \quad g_{SM} \cos{\theta_W} = g_5 \int_{\ell_0}^{\ell_1} \frac{\ell_0^4}{z^4}\phi^2_{f_L}(z)\phi_{L_3}(z) dz, 
\end{equation}
where $g_{SM}, \cos{\theta_W}$ are the 4D coupling and weak mixing angle. The $z$-dependent functions $\phi_{f_L},\phi_W,$ and $\phi_{L3}$ are the extra dimensional profiles of the left-handed SM fermions, the $W^{\pm}$ and $W_3$, respectively. As it involves $\phi_{f_L}$, this matching necessarily introduces a dependency on the fermion masses and boundary conditions into the gauge boson normalization.

Having normalized the profiles by matching charged current coupling, rather than by setting the norm of the profiles to 1, the 4D gauge fields have non-canonical normalization:
\begin{equation}
\label{eq:SMcoupl}
-\frac{1}{4g^2_5 }\int_{\ell_0}^{\ell_1}\frac{\ell_0}{z}(L_{\mu\nu})^2 dz \cong -\frac{1}{4 g^2_4}\Big(\frac{g^2_4}{g^2_5 }\int_{\ell_0}^{\ell_1}\frac{\ell_0}{z}\phi_W(z)^2 dz \Big)W_{\mu} \Box W^{\nu} \equiv -\frac{1}{4 g^2_4} \Pi_{WW} W_{\mu} \Box W^{\nu},  
\end{equation}
and $\Pi_{WW} \ne 1$. Because the matching conditions depend on the fermions' position in the extra dimension, $\Pi_{WW}$ inherits a sensitivity to the fermion profile. For example, the fermions could simply be affixed to the UV brane, in which case their profile is $\phi_f(z) \sim \delta (z - \ell_0)$. The CFT interpretation of these fermions is as fundamental states probing the strongly interacting sector~\cite{Contino:2004vy}. More realistic, massive fermions must  feel the effects of EWSB, thus their profiles must be spread out into the extra dimension. The parameter which controls the fermion wave functions is a 5D Dirac mass term:
\begin{equation}
\label{eq:5dmass}
M \, \bar{\psi} \,  \psi \ .
\end{equation}
A mass term like (\ref{eq:5dmass}) can be introduced for every SM chiral fermion. The mass parameters are usually expressed in units of the AdS curvature $k \sim \frac{1}{\ell_0}$,  $c_{L,R} = M_{L,R}\ell_0$, for left and right-handed fermions, respectively. For $c_L > 1/2$, the left-handed fermion zero-mode is located near the UV brane $\ell_0$ and its interpretation from the dual side is a fermion with a small admixture with the CFT sector. Similarly, left-handed fermions with bulk mass $c_L < 1/2$ are located near the IR brane, and therefore mix with the CFT substantially~\footnote{For a right-handed fermion $c_R < -1/2$ corresponds to a fundamental fermion, while $c_R > -1/2$ corresponds to a fermion with a large CFT admixture }. For the intermediate mass case $c_L =1/2$, also called a `conformal mass', the coupling of the elementary fermion to the CFT is marginal. If any of the SM fermions are pushed into the bulk, we must also extend color $SU(3)$ to a 5D gauge symmetry. KK decomposition on these fields yields a tower of gluonic excitations $G_{KK,i}$.

With the above matching scheme, $\Pi_{WW}, \Pi_{ZZ}$ contain all the effects of the new physics on the EW sector, and the precision electroweak observables (PEW) can be calculated from $\Pi_{WW}, \Pi_{ZZ}$ according to the usual techniques~\cite{Peskin:1990zt, Golden:1990ig,Holdom:1990tc, Burgess:1993mg, Burgess:1993vc}. Assuming the fermions to be localized on the UV brane ($z = \ell_0$), the result for $S$ to leading order in $(\log{\ell_1/\ell_0})^{-1}$ is:
\begin{equation}
\label{eq:ESS}
S \sim \frac{6\pi}{g^2_4 \log{\ell_1/\ell_0}},
\end{equation}
which is positive and typically $\mathcal O(1)$~\cite{Agashe:2003zs, Csaki:2003zu, Cacciapaglia:2004jz} -- far away from the preferred PEW value $S \lesssim 0.1$~\cite{Amsler:2008zz}.  On the other hand, if the left-handed fermions were localized near the IR brane ($c_L \ll 1/2$), as in the original proposal by Randall and Sundrum, the total contributions to the $S$ parameter would be negative and large~\cite{Csaki:2002gy}, again in contradiction with experimental constraints. 

\section{The Cure: Mass-Matching}
\label{sec:massmatch}

From the equations (\ref{eq:match}, \ref{eq:SMcoupl}) we can identify the ingredients which enter into the calculation of the PEW parameters: the gauge boson profiles, the left-handed fermion profiles and the geometry where they are computed. The requirements of the electroweak symmetry breaking boundary conditions do not leave much flexibility in the gauge boson profiles. We can, however, modify the bulk fermion profile. As mentioned earlier, changing the fermion profiles means altering the bulk fermion masses, $c_L$ and $c_R$. The estimate of $S$ in~(\ref{eq:ESS}) assumed UV located left-handed fermions, or $c_L \gg 1/2$. This is an obvious starting point for the light quarks, as it describes fundamental fermions interacting with, but external to, a strong near-conformal sector.  The smaller $c_L$ is, the more the left-handed fermions are pushed into the bulk of the extra dimension, and the more the SM quarks become an admixture of elementary quarks and baryonic bound states of the new strong interaction.


One can use light fermion compositeness to suppress contributions to the $S$ parameter. To see why this is the case, let us first argue in the case of unbroken electroweak symmetry. With EWSB shut off, zero-mode, massless gauge bosons are flat in the bulk, and they are orthogonal to their KK excitations. By setting $c_L \sim 1/2$, the profile of the zero-mode fermion current $\bar{\phi}_{f_L} \gamma^{\mu}\phi_{f_L} (z)$ is also flat, and therefore the fermion current is also orthogonal to the KK spin-1 resonances. By tuning the fermions such that they decouple from the KK gauge bosons, there are no new, tree-level contributions to the electroweak sector, and therefore $S \cong 0$.  Once EWSB-effects are included, the KK gauge boson profiles are no longer perfectly flat. We can still tune the fermion profiles to cause a cancellation in $S$, however the cancellation will now occur with a $c_L$ value slightly different from $1/2$.  Ideally, one would hope for a symmetry to enforce $c_L \sim 1/2$. Unfortunately no such symmetry has been found, so $c_L \sim 1/2$ and thus $S \cong 0$ can only be achieved via tuning. 

Within the experimental uncertainty on the PEW parameters, a range of $c_L$ values is permitted for a given value of the UV cutoff. The value of $c_L$ which leads to $|S|<0.5$ is shown below in Fig.~\ref{fig:Sallowed}.  In Fig. ~\ref{fig:massesandcouplings} and \ref{fig:couplings2} we plot the same region of $c_L$ and $l_0$, overlying the values of resonance mass $W_{KK}$ and their couplings to the light fermions, respectively. The numbers quoted in Fig.~\ref{fig:couplings2} are $\kappa=g_{W_{KK} f_L f_L}/g_{SM}$, the ratio of SM fermion coupling to first charged KK mode versus the SM $W^{\pm}$ boson. The zero-mode fermion coupling to the neutral KK resonances has a similar size, though the exact value depends on the fermion's charge and hypercharge. 
  \begin{figure}[h!]
 \centering
 \begin{minipage}[c]{0.3\textwidth}
 	\centering
	\includegraphics[width=2.1in , height=2.5in]{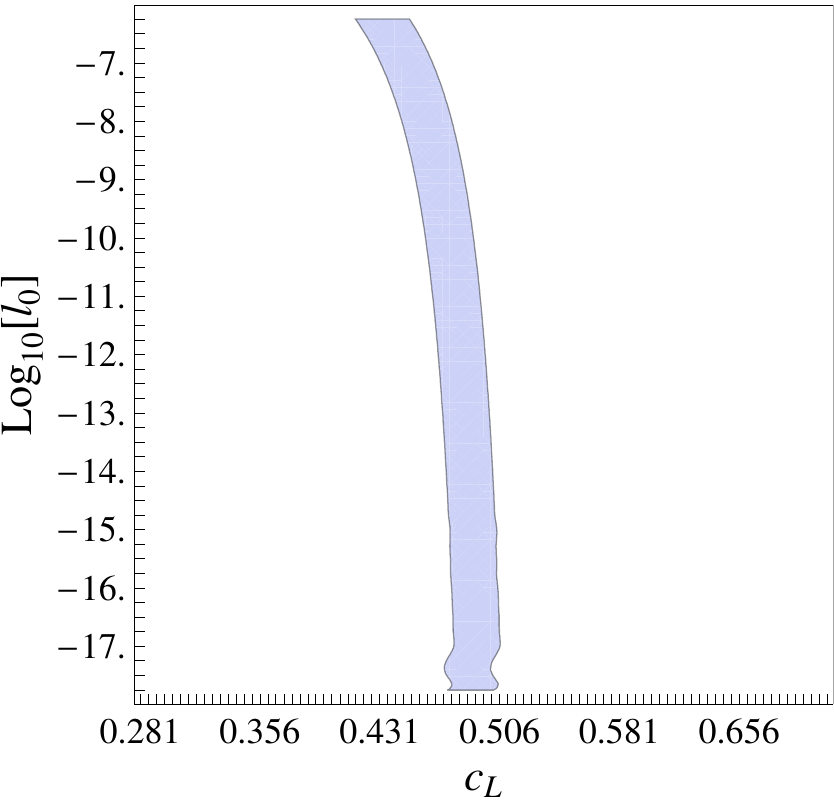}
 	\caption{$|S|<0.5$ region vs. $c_L$ and $\ell_0$.}
 	\label{fig:Sallowed}
 \end{minipage}
 \hspace{0.55cm}
 \begin{minipage}[c]{0.3\textwidth}
 	\centering
	\includegraphics[width=2.1in, height=2.5in]{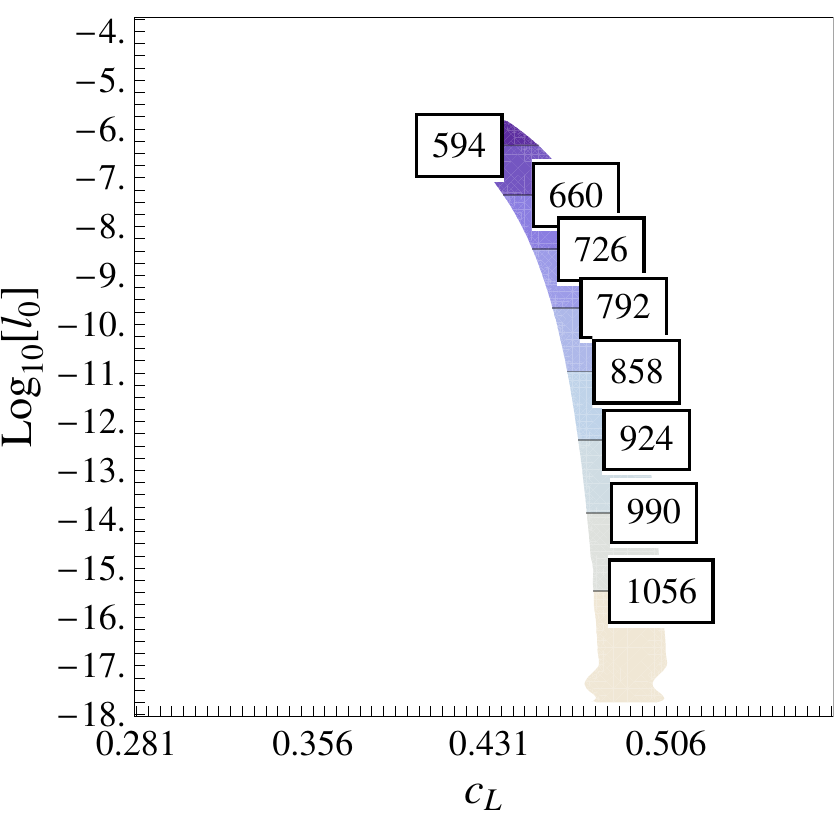}
 	\caption{Masses with $|S|<0.5$ and $|T|<0.3$ }
	\label{fig:massesandcouplings}
\end{minipage}
\hspace{0.55cm}
\begin{minipage}[c]{0.3\textwidth}
\centering
	\includegraphics[width=2.1in, height=2.5in]{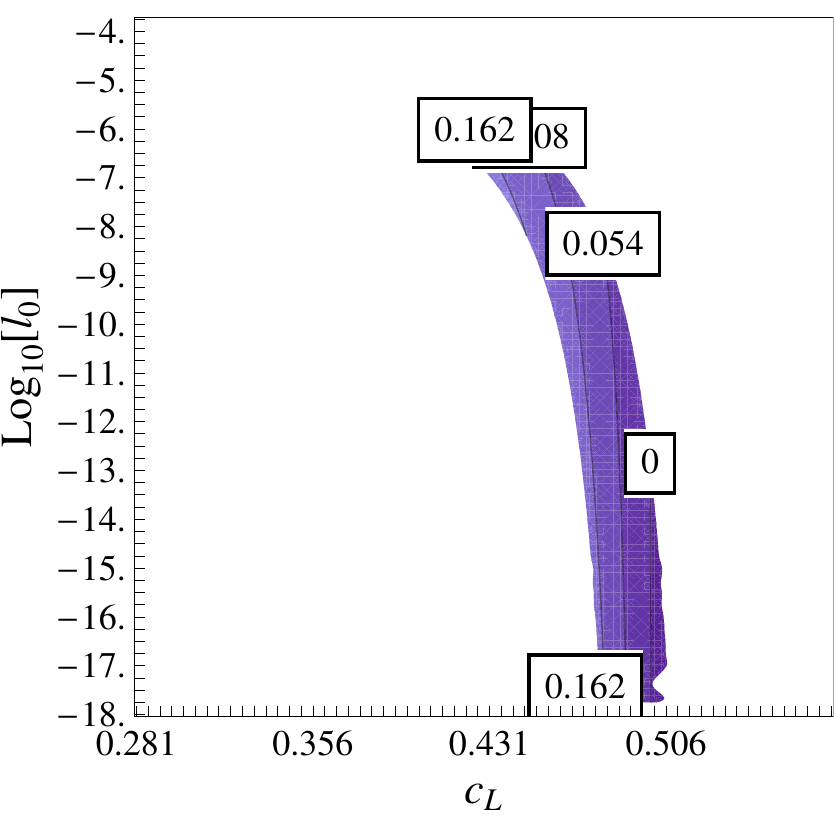}
	\caption{Couplings with $|S|<0.5$ and $|T|<0.3$ }
	\label{fig:couplings2}
\end{minipage}
\end{figure} 
   
Notice that varying the UV cutoff has little effect on $S$, so the same narrow range of $c_L$ values is allowed over a wide range of $\ell_0$. On the other hand, the value of $\ell_0$ does control the separation between the $W$ and $W_{KK}$ masses, see center Fig.~\ref{fig:massesandcouplings}. The $T$ parameter constraint $|T| < 0.3$ is responsible for cutting off the rightmost two figures at $\text{log}(\ell_0) \leqslant -6,\ M_{W_{KK}} \geqslant 600 \ \gev$. Stronger bounds on $T$ will push up the resonance masses even further. For example, $T<0.1$ implies the $W_{KK}$ mass should be heavier than 730 GeV. Similarly, lowering the bound on $S$ squeezes the allowed $c_L$ region even narrower. Figures~ \ref{fig:massesandcouplings}-\ref{fig:couplings2} have been generated assuming massless zero-mode fermions. Masses for the first and second generation quarks and leptons, the only fermions which play a role in the extraction of $S, T,$ etc., are negligible and do not modify our results substantially. 

Once the values of $c_L$ and $l_0$ are set to achieve a desired $S$, the value of the charged current coupling of W is fixed by Eq.~(\ref{eq:SMcoupl}). As we have chosen to match the 5D theory to the 4D theory through the charged current coupling, two consequences follow. First, the triboson coupling $W^+W^-Z^0$, which is determined by the overlap of the  $\phi_{W^-} ,\phi_{W^+}$  and $\phi_Z$ profiles, is not necessarily equal to is SM value, $g_{SM} \cos{\theta_W}$ -- so we must check that there are no large anomalous triboson couplings. Second, the simplest choice for the right-handed fermion mass terms $c_R = -c_L$ -- left and right-handed fermions localized in the same position of the extra dimension -- leads to unacceptable values for the right-hand quark couplings to the $Z^0$ boson. Both of these consequences can be dealt with without changing the mass-matching. In the region bounded by $S$ and $T$, we find all anomalous trilinear couplings are small and LEP bounds~\cite{Amsler:2008zz} do not reduce the parameter space any further. Also, there is enough freedom in the $c_R$ choice to avoid constraints from couplings to right-handed fermions, though this does come at the expense of a loss of universality (different $c_R$ for different flavors and generations) making flavor-constraints non-trivial~\cite{Cacciapaglia:2004rb}.

The left-handed fermion $c_L$ value clearly has ramifications on the mass of the fermionic KK excitations. This can be seen in Fig.~\ref{fig:mm}: whenever $c_L\sim1/2$, the mass of the KK fermions and KK gauge bosons are nearly equal. The observability of the mass-matched spectrum and its consequences in the resonance couplings,  a `smoking-gun' signal for the Cured Higgsless class of models, will be our focus for the rest of this paper.  One final observation which will have important phenomenological implications later on is that the couplings of the KK gauge bosons to the SM fermions in the allowed $S$ window is small but non-zero, due to EWSB effects.

 A similar mass-matching occurs if one chooses to alter the geometry of the extra dimension, rather than the fermion profiles -- a scenario known as Holographic Technicolor. In this case, the mass-matching occurs between two different spin-1 particles,  the vector and axial combinations of the $SU(2)_L$ and $SU(2)_R$ KK gauge bosons, rather than between particles of different spin. The reason $S \cong 0$ in Holographic Technicolor can easily be seen by recasting the definition of $S$ as a sum over narrow resonances:
\begin{equation}
S = 4\pi \sum_i \Big(\frac{F^2_{Vi}}{M^2_{Vi}}- \frac{F^2_{Ai}}{M^2_{Ai}} \Big), 
\end{equation}
where $i$ runs over all resonances. Clearly, the more degenerate the vector and axial resonances are, the smaller $S$ becomes~\footnote{The deformations from AdS geometry required for a near-degenerate spectrum and small $S$ must be larger than NDA estimates~\cite{Agashe:2007mc}. Therefore, like Cured Higgsless, Holographic Technicolor achieves small $S$ only through fine-tuning,}. It is interesting that, despite the differences in how $S \cong 0$ is achieved, both Cured Higgsless and Holographic Technicolor rely on mass-matched spectra. The signatures of Holographic Technicolor, both at low and high luminosity, were explored in Ref.~\cite{Hirn:2007we, Hirn:2008tc}.

Another interesting fact about mass-mathing is it allows one to distinguish between the continuum 5D Higgsless models and the current (4D) deconstructed models with `ideal fermion delocalization'~\cite{Chivukula:2005bn,Chivukula:2005xm, Chivukula:2006cg}. The current deconstructed models also achieve $S \sim 0$ by fine-tuning the admixture of underlying fermions which composes the SM fermions. However, where 5D models have the constraint that all fields must feel the same extra dimension, deconstructed models do not. 
As a result, although the particular admixture of underlying fermions composing the SM must be matched to the admixture of underlying gauge bosons composing the $W^{\pm}, Z$, their overall mass scales are not related in a deconstructed model. In fact, with the flavor implementation of the current models, the fermion mixture orthogonal to the SM is forced to be much heavier than the extra, massive gauge bosons. Another important phenomenological distinction is that deconstructed models in Ref.~\cite{Chivukula:2005bn,Chivukula:2005xm, Chivukula:2006cg} do not contain a KK gluon.

 In the next section we briefly explore the remaining parameter constraints on mass-matched Higgsless scenarios. After honing the parameter space, we then investigate several distinct signatures of mass-matched models and how they can best be identified.
 
 \section{Collider Constraints}
\label {sec:hlessPEW}
 
Within a mass-matched scenario, the constraint from the $S$ parameter is under control, as are anomalous triple-gauge-boson couplings. The size of the extra dimension, and therefore the allowed KK gauge boson mass, is constrained by the $T$ parameter. However, one may worry that the bounds on new light resonances from  LEP or Tevatron studies are more severe. Of all direct bounds from LEP and the Tevatron, a search for signatures containing a high transverse momentum $Z^0$ boson in CDF-II data are the most stringent~\cite{CDFnote}. Other bounds, such as compositeness~\cite{hep-ph/0106251, Amsler:2008zz}, high mass resonances into $e^+ e^- , \,  e^{\pm} \,\nu$~\cite{:2007sb, arXiv:0707.2524, hep-ex/0611022} and searches of resonances into $W^{\pm}\, Z^0$ and $W^+\, W^-$\cite{CDFnote2}, turn out to be less restrictive than the high-$p_T$ objects search for our study. 

The easiest limit from the high-$p_T$ study to apply to our scenario is from the exclusive channel $pp \rightarrow Z(\ell^+\ell^-) + \ell' + \slashchar{E}_T$, where the leptons are muons or electrons. Within the Higgsless setup, the primary source of high-$p_T$ $Z^0$ bosons in this exclusive channel is the production of a charged $W_{KK, i}$ which subsequently decays into $W^{\pm}Z^0$.  High-$p_T$ $Z^0s$ can also come from the decays of KK quarks, however these decays always involve extra hard jets. Allowing extra jets into the analysis lets in additional large SM backgrounds, such as $Z + \jets$. Even with larger SM backgrounds, constraints from high-$p_T + Z^0 + \jets$ may be  important, however the relevant limits cannot be extracted from the only high-$p_T Z^0 + X$ study we know about~\cite{CDFnote}.

\begin{figure}[h!]
 \centering
 \begin{minipage}[c]{0.45\textwidth}
	 \centering
   	\includegraphics[width=2.75in, height=2.05in]{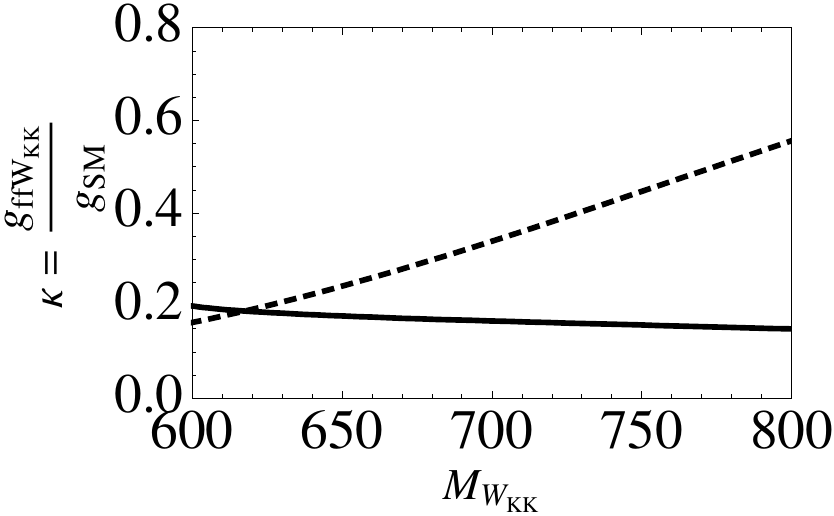}
	\caption{Combined $S,T$ constraints (solid line) and the Tevatron bound from high-$p_T$ objects (dotted line)}.
	\label{fig:constraints1}
\end{minipage}
\hspace{0.2in}
\begin{minipage}[c]{0.45\textwidth}
\centering
	\includegraphics[width=2.75in]{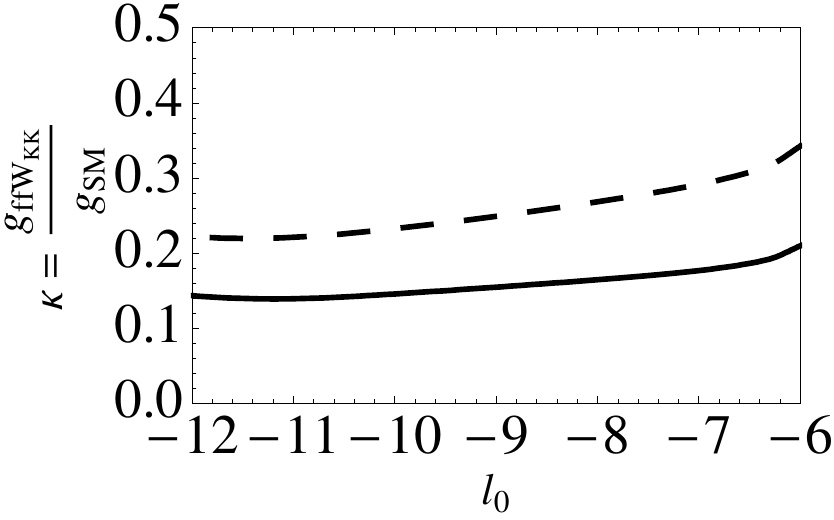}
	\caption{The couplings of left-handed zero-mode fermions to the first (solid line) and second (dashed line) KK charged gauge boson resonances as a function of $\ell_0$ for the same points as the left-hand plot}
	\label{fig:constraints2}
\end{minipage}
 \end{figure}

 Restricted to $s$-channel charged resonance production, the high-$p_T$ study sets an upper bound on the coupling between a KK gauge boson of a given mass and SM (zero-mode) fermions.  We estimated the high-$p_T$ constraint by repeating the analysis of~\cite{CDFnote} on the the generated signal -- albeit with poorer Monte Carlo tools -- and requiring fewer than $3~Z^0(\ell^+\ell^-)$ bosons with $p_T > 100\ \gev$ after all cuts. The region of allowed coupling from the high-$p_T$ bound, along with the bound on the gauge KK resonance coupling to light fermions from ($S$,$T$) are shown in Figure~\ref{fig:constraints1}. In Fig.~\ref{fig:constraints1}, the coupling of the first and second tier of resonances are denoted as fractions of the SM $SU(2)_W$ coupling:~$\kappa_1\cdot g_{SM}$ and $\kappa_2 \cdot g_{SM}$,  and we place limits on $\kappa$. We see that the strongest bounds come from indirect  ($S$ and $T$)  constraints except when the resonances are lighter than $620\ \gev$.
 
Two conclusions are to be drawn from these plot: 1) Bounds still allow for sizable couplings to light fermions, therefore s-channel production of charged and neutral KK gauge boson resonance is their dominant  production mechanism, 2.) The second tier of resonances have a {\em stronger} coupling to light fermions than the first tier. 

Finally, the third generation in Cured Higgsless models must be treated specially due to the large top quark mass. To get an acceptable top quark mass, the 3rd generation quarks must be localized closer to the IR and are therefore more strongly coupled to $W_{KK}$\@. While direct bounds of resonance decays to third generation quarks, $W' \rightarrow t \, b$ are not severe~\cite{Aaltonen:2009qu}, there is a tension between realistic top mass and accurately measured $Z \bar{b}_L b_L$ couplings~\cite{Amsler:2008zz}. One possible solution is to change the representation of the third quark generation under the bulk gauge group~\cite{Agashe:2006at}. Embedding the left-handed $(t_L, b_L)$ doublet within a $(2,2)$ of $SU(2)_L \otimes SU(2)_R$, one immediate consequence is an extra fermion, $X$\@. The $X$ fermion is colored, has elctromagnetic charge $+\frac{5}{3}$, and a very interesting phenomenology of its own. The solution of Ref.~\cite{Agashe:2006at}  has been implemented in several other EWSB scenarios, such as Little Higgs\cite{Katz:2005au} and Composite Higgs scenarios \cite{Agashe:2005dk, Contino:2006qr}. Some work has already been done to characterize the phenomenology of this new fermionic sector\cite{Cacciapaglia:2006gp, Contino:2008hi, Brooijmans:2008se}. As the peculiarities of the third generation are not essential for solving the $S$ parameter problem with Higgless models we will not discuss it in this paper. 

A subset of the resulting spectrum for a Cured Higgsless scenario with $c_L = 0.446, \ell_0 = 10^{-8}\ \gev^{-1}$ -- a point which complies with all precision electroweak and LEP/Tevatron constraints -- is shown below in Fig.~\ref{fig:spect}. The complete parameter specifications are deferred to Appendix (\ref{CHLPAR}).
\begin{figure}[h!]
\centering
\includegraphics[width=3.6in, height=2.0in]{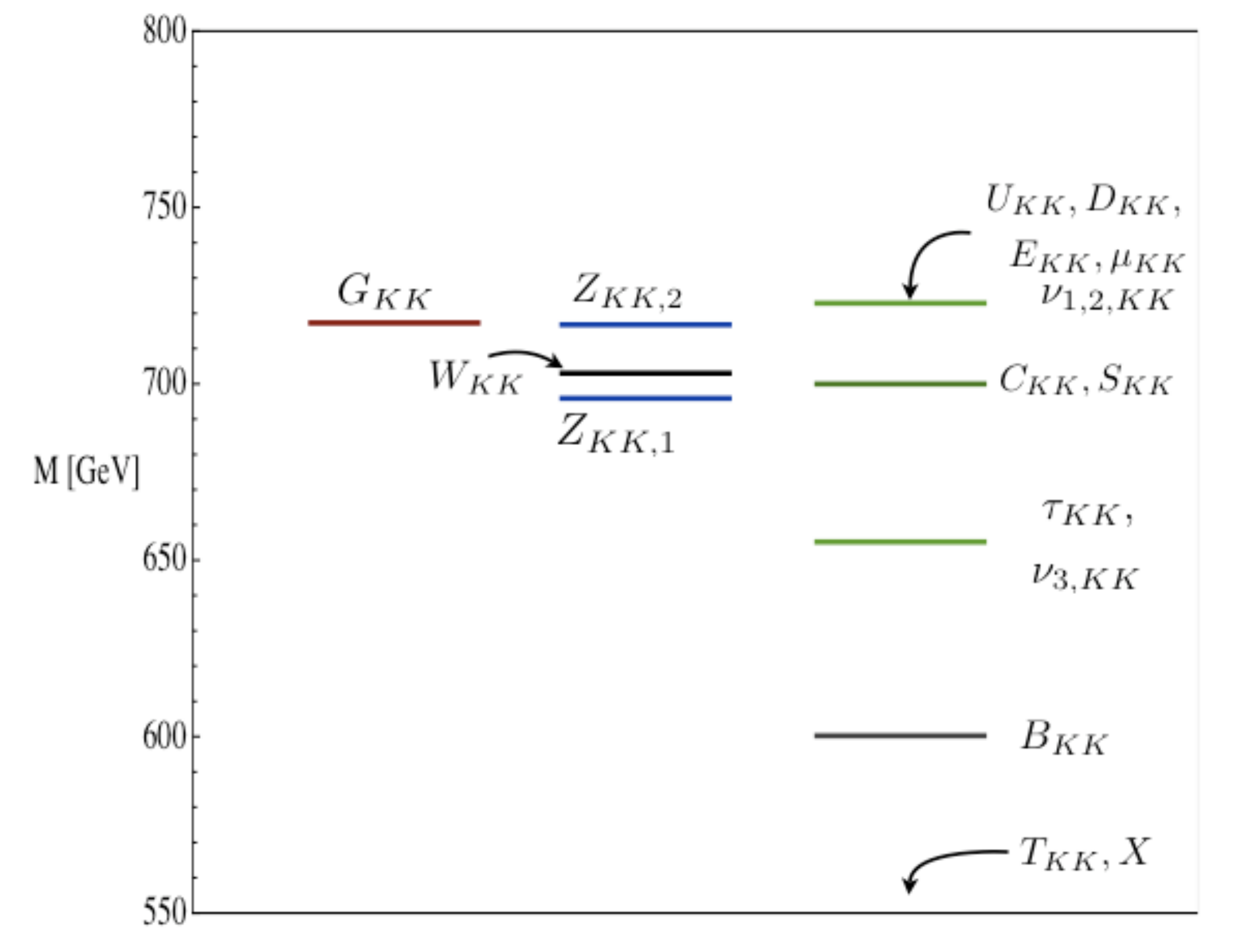}
\caption{ Masses of the first KK modes in a Cured Higgsless scenario with  $c_L = 0.446, \ell_0 = 10^{-8}\ \gev^{-1}$. The $c_R$ of the first and second generation fermions have been chosen to reproduce the SM right handed fermion couplings to the $Z^0$, and the third generation quarks have been implemented using the alternative $SU(2)_L \otimes SU(2)_R$ representation suggested in Ref.~\cite{Agashe:2006at}. The full set of parameters for this point can be found in Appendix (\ref{CHLPAR}). }
\label{fig:spect}
\end{figure}
Most of the KK resonances have comparable masses, for example $M_{U_{KK}} \sim M_{Z_{KK,1}}\sim M_{G_{KK}}$, where $G_{KK}$ is the KK excitation of the gluon and $Z_{KK, 1}$ is the KK photon.  The KK excitations from the third generation quarks are split from their first and second generation fellows because the localization of the zero-mode is not flat, but peaked towards the IR to accommodate the top mass.  

  
\section{Identifying the Mass-Matching mechanism}
\label{sec:hlessPHENO}

Having solidified the acceptable range of parameters in mass-matched Higgsless models, we now move on to identifying and discussing the disctinct features of these models and how they will manifest in LHC observables.

The most obvious feature is the matching in the spectrum of baryons (KK fermions) and mesons (KK gauge bosons), however 
mass-matching has other phenomenological repercussions. In the original warped Higgsless scenarios, light fermions are located near the UV brane and are therefore mostly elementary. Gauge boson KK resonances (or composites) are localized near the IR brane. As there is little overlap, the coupling between light fermions and resonances is small. More concretely, the coupling is approximately suppressed by
\begin{equation}
\lambda_{l,h^2}\sim\left(\frac{l_0}{l_1}\right)^{2 c_L -1} \ .
\end{equation}
where $\lambda_{l, h^{2}}$ is the trilinear among one light fermions and two heavy states and $c_L \gg 1/2$ for an elementary left-handed fermion. On the other hand, mass-matching occurs when the fermion zero-modes are almost flat in the extra-dimension and their overlap with two KK resonances is just suppressed by a volume factor. This results in a non-negligible coupling, 
\begin{equation}
\lambda_{l,h^2}\sim\frac{1}{\log\left(\frac{l_0}{l_1}\right)} \ .
\end{equation}  

\begin{figure}[!h]
\begin{minipage}[c]{0.45\textwidth}
	\centering
	\includegraphics[width=3.0in, height = 2.0in]{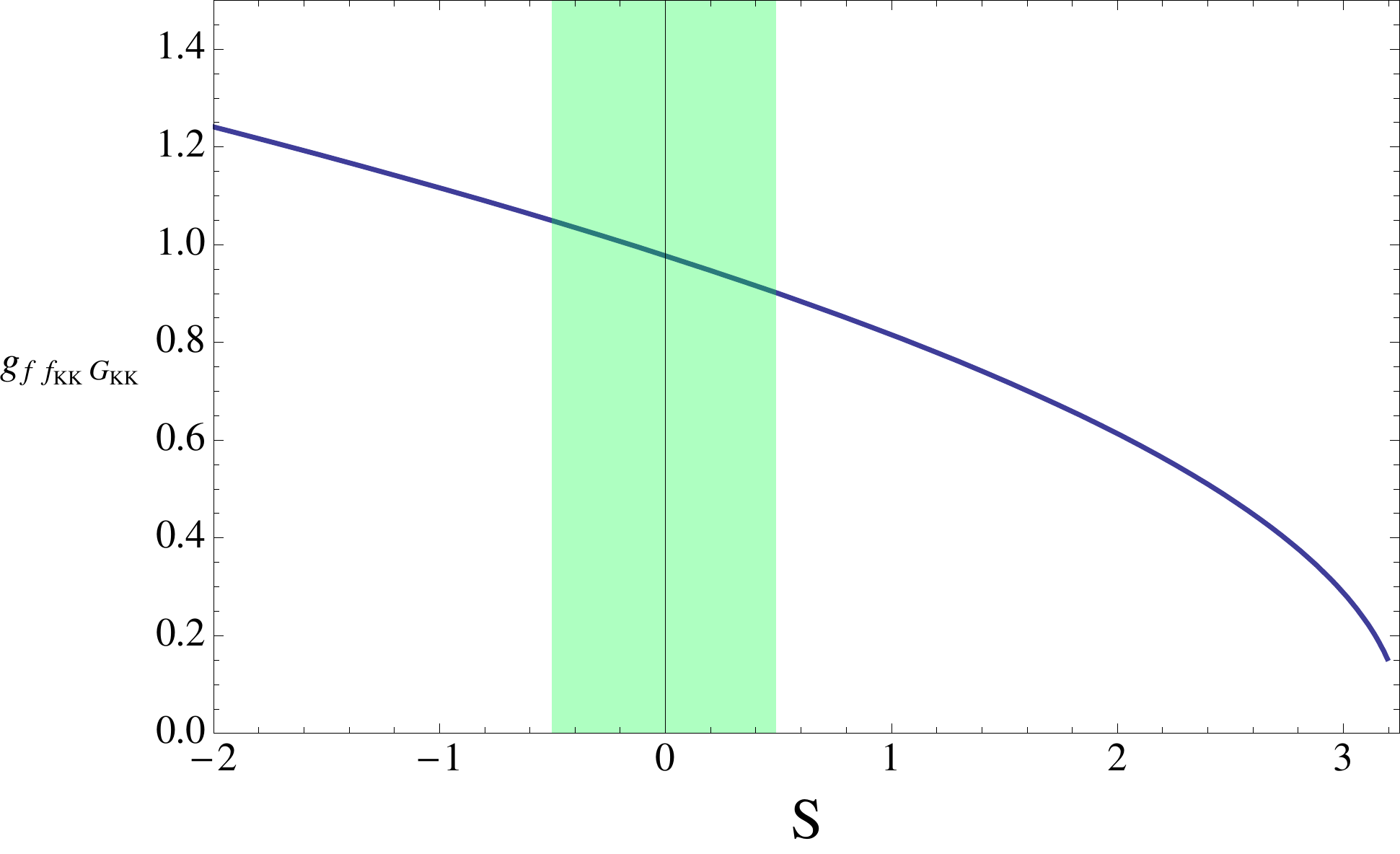}
	\caption{KK fermion coupling to KK gluon and light fermion as a function of the value of $S$. The shaded region indicates $|S|< 0.5$.}
	\label{fig:gKKvsS}
\end{minipage}
\hspace{0.3in}
\begin{minipage}[c]{0.45\textwidth}
\centering
	\includegraphics[width=3.0in, height = 2.0in]{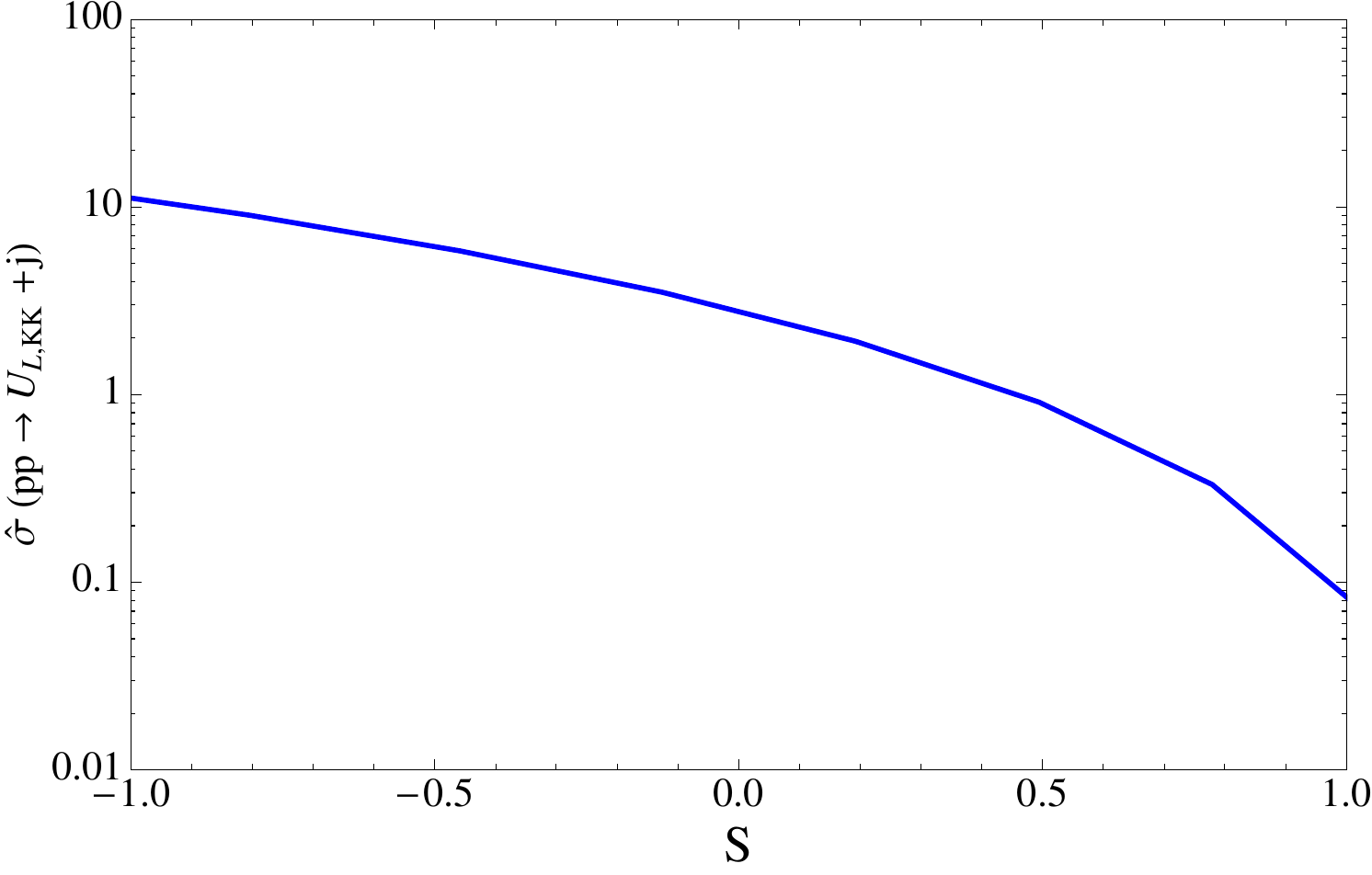}
	\caption{Partonic cross section for $pp \rightarrow j + U_{L,KK}$ at center of mass energy $10\ \tev$ as a function of $S$.}
	\label{fig:xsecS}
\end{minipage}
\end{figure}

To demonstrate this effect, in Fig.~\ref{fig:gKKvsS} we plot the KK fermion-KK gluon-light fermion coupling versus the value of $S$\@. Negative values of $S$ occur when the light fermions are fully composite~\cite{Csaki:2002gy} (left-hand side of Fig.~\ref{fig:gKKvsS}), while elementary light  fermions (right-hand side of Fig.~\ref{fig:gKKvsS}) correspond to a large and positive value for $S$~\cite{Barbieri:2003pr}. The results from LEP indicate one should live very precisely in between -- where SM fermions are an admixture of composite and elementary~\cite{Cacciapaglia:2004rb} and $\lambda \sim {\cal O} (1)$. Thus, assuming some degree of compositeness in the light fermions is concurrent with a large trilinear coupling among light fermions and two KK resonances.

Using the coupling in Fig.~\ref{fig:gKKvsS}  we can compute the partonic cross section to singly produce a $U_{L,KK}$ in a $pp$ collision. We use the approximation of massless fermions, compute using the CTEQ5L~\cite{Lai:1999wy} parton distribution functions, and take $\sqrt s = 14\ \tev$. The partonic cross section,  shown in Fig.~\ref{fig:xsecS} above, clearly follows the same trend as the KK fermion - KK gluon - zero-mode fermion coupling; the cross section at $S \simeq 0$ is nearly two orders of magnitude larger than when the fermions are mostly fundamental ($c_L \gg 1/2, S \gg 0$). However, the behavior of the cross section is slightly more complicated than what is depicted in Fig.~\ref{fig:xsecS}. When $c_L$ is exactly 1/2, the KK fermion - KK gluon - zero-mode fermion coupling is large, but the coupling of the KK gluon to {\em two} zero-mode fermions vanishes by the same orthogonality argument given in section~\ref{sec:massmatch}. Therefore, if the cancellation in $S$ required $c_L = 1/2$, then the single KK fermion signal would be hopelessly small. However, the $W/Z$ wave functions are not quite flat because of EWSB, so $c_L$ slightly less than $1/2$ are preferred, and we recover a large single-$U_{KK}$ rate. As we go to smaller $\ell_0$, the resonances become heavier and EWSB has a smaller effect on the wave functions. This moves the preferred $c_L$ region closer to $1/2$. For such heavy resonances it is therefore possible to have mass matching without a large single KK fermion cross section. Such an unlucky scenario would still have mass-matching, it would just be harder to find.

Summarizing, identifying the mass-matching mechanism relies on specifying two characteristic features
\begin{enumerate}
  \item A mass degeneracy between the gauge boson resonances and the fermionic resonances. We will extract the value of the neutral gauge boson KK ($M_{Z_{KK}}$) via  in the dilepton channel $pp \rightarrow Z_{KK, i} \rightarrow \ell^+\ell^-$. To find the mass of the fermion KK, we explore the clean pair production channel $pp \rightarrow Q_{KK}\, Q_{KK} \rightarrow 4\ell+ 2\ \jet$ channel.  
    \item Large trilinear couplings among gluon and fermion KK resonances and light fermions. Single and pair production of the KK fermion will be the cleanest channels to probe this coupling.  
\end{enumerate}


\subsection{Simulation basics}
\label{sec:simbasic}

To more thoroughly investigate the above concepts, we choose a point in the CHL parameter space consistent with existing bounds and explore its phenomenology. The point we choose is exactly the one depicted in Figure~\ref{fig:spect}. The 5D Lagrangian parameters which generate this point are displayed in Table \ref{eq:CHLPOINT}.
\begin{gather}
\ell_0 = 10^{-8}\ \gev^{-1},\quad c_L = 0.446,  \nn \\ 
c_{\ell_R} = -0.68, \quad c_{q_{R_1}} = c_{q_{R_2}} = -0.6 \nn \\
c_{q_{L_3}} = 0.26,\quad c_{b_{R_3}} = -0.73,\quad c_{t_{R_3}} = -0.42,
\label{eq:CHLPOINT}
\end{gather}
and the masses of the KK states corresponding to these parameters are given in Appendix~(\ref{CHLPAR}). These particular parameters were chosen following the strategy mentioned in section~\ref{sec:massmatch}. For the third generation, which plays no role in this paper except through its influence on resonance widths, the $c_{L,R}$ parameters are exactly those in Ref.~\cite{Cacciapaglia:2006gp}. We chose this point as a benchmark as it has rather low-mass resonances, an essential feature of the Higgsless mechanism of perturbative $W_L\,W_L \rightarrow W_L W_L$ unitarization. As the resonances become heavier, they become more strongly interacting and the Higgsless paradigm becomes indistinguishable from a traditional technicolor model.

To analyze the collider signatures at this and similar points, we used the following software tools: We generated the events at the parton level with the Monte Carlo generator MadGraphV4.4~\cite{Alwall:2007st} with a center of mass energy of 14 TeV. For the processes we are looking at, the change from 14 TeV to 10 TeV results in a reduction of about a factor of $2$. We used BRIDGE~\cite{hep-ph/0703031} for particle decays, PYTHIAv6.4~\cite{Sjostrand:2006za} to include showering-hadronization effects and PGSv4~\cite{PGS} to estimate detector effects. All of the events were generated with the following basic, parton-level cuts:
\begin{itemize}
  \item leptons with $|\eta|<$ 2.5, $p_T> $10  GeV and parton level isolation cuts  $\Delta R_{\ell\ell}>$0.4,  $\Delta R_{\ell j}>$0.4
  \item jets with   $|\eta|<$ 2.5, $p_T> $15  GeV and isolation cuts  $\Delta R_{jj}>$0.4
\label{itemi1}
\end{itemize}
 To parameterize the detector response, the PGS parameters provided in {\tt
  pgs$\_$card$\_$ATLAS.dat} were used.

 
\subsection{Single production fermion KK: the $2\,\ell+ 2\ \jet$ channel}
\label{ssec:singleKK}

\begin{figure}[t!]
\begin{center}
\centering
\begin{minipage}[c]{0.45\textwidth}
	\centering
	 \includegraphics[width=2.25in, height = 1.8in]{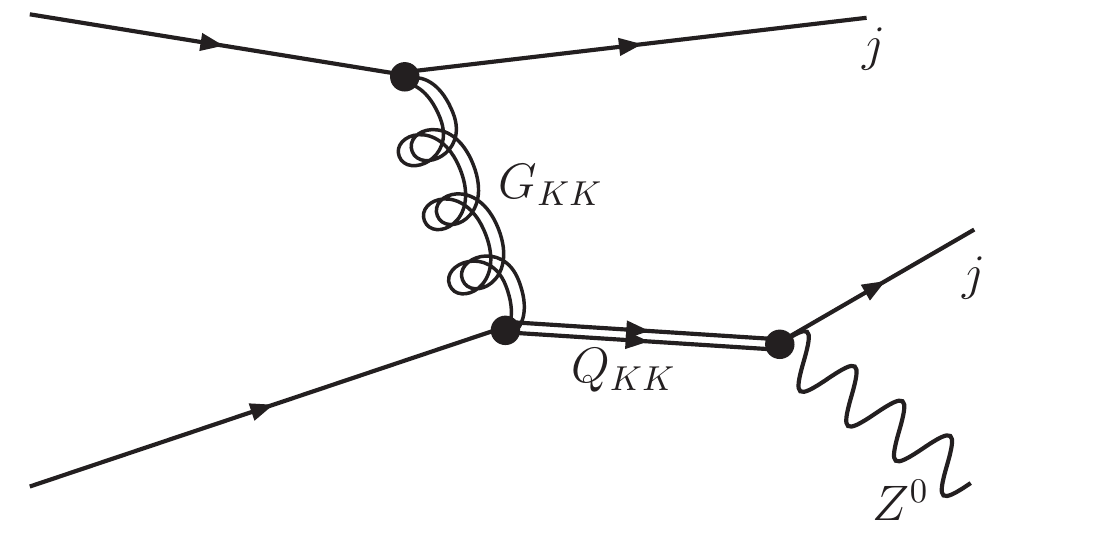}
	  \caption{Single production KK fermion diagram}
	   \label{singlediag}	
\end{minipage}
\hspace{0.2in}
\begin{minipage}[c]{0.45\textwidth}
	\centering
 	\includegraphics[width=1.9in,height = 2.75in, angle=90]{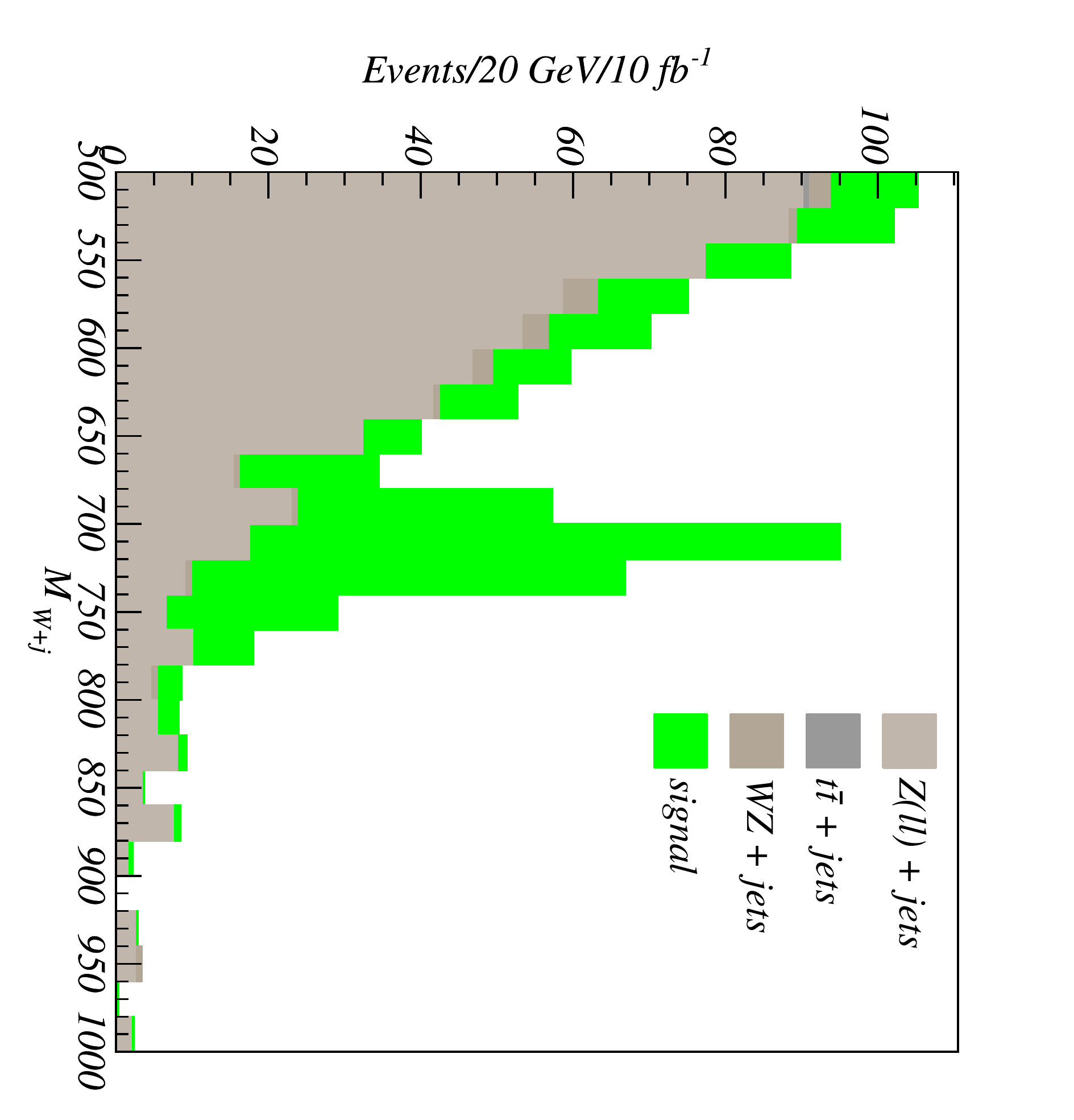}
	\caption{Single production invariant mass reconstruction in the 2 $\ell$ + 2 $j$ channel.}
	\label{fig:singKKmass}
\end{minipage}
 \end{center}
 \end{figure}

Haven chosen a benchmark point, the first signal we investigate is the single production of a KK fermion with a light jet. Single fermion KK resonances are produced predominantly by a exchange of a KK gluon as depicted in Fig.~\ref{singlediag}. The total cross section is proportional to 
\begin{equation}
\sigma_{tot} \propto \lambda^2_{l, h^2} \, \lambda^2_{l^2, h}
\end{equation}
where $\lambda_{l^{1(2)}, h^{2(1)}}$ is the trilinear among one (two) light fermions and two (one) heavy states. As explained in Section \ref{sec:hlessPHENO}, in usual warped models those trilinears would be very suppressed, whereas the mass-matching mechanism relies on a partly-composite quark, and therefore sizable $\lambda$.  
To probe those couplings $\lambda$ we looked at the process shown in Fig.~\ref{singlediag} where $Q_{KK}$ denotes first and second generation KK quarks,
\begin{equation}
Q_{KK} \,  =  \, U_{KK} \ , \, D_{KK} \ , \, S_{KK}\ , \, C_{KK} \ .
\end{equation}
Since single production requires an initial state quark, the contribution from the third generation quark KK states (including the $X$) is negligible. Similarly, due to the larger $u$ and $d$ parton distribution functions, the first generation KK states completely dominate the signal.

Because the mass of the KK gluon is so close to the mass of the KK quark, decays $Q_{KK} \rightarrow G_{KK} + \jet$ are kinematically suppressed. As a result, the KK quarks are quite narrow and decay predominantly into $W^{\pm}+\jet$ or $Z^0 + \jet$. Of the subsequent $W^{\pm}, Z^0$ decays, the leptonic $Z^0$ decays are the cleanest and have the highest potential for discovery, thus the final state we consider is $2 \ell $ + $2\ \text{light jets}$, where $\ell = e, \mu$. 

The Standard Model backgrounds to this process are primarily 1.)  leptonic $t \, \bar{t} + \jets$, 2.) $W^{\pm} \, Z^0$+jets and 3.) $Z(\ell^+\ell^-)+\jets$. To reduce the sizable background, we impose the following cuts on the reconstructed events:
\begin{enumerate}
  \item $n_{\ell}=2$, same-flavor, opposite sign leptons and $m_{\ell \ell} = m_Z \pm 20$ GeV
  \item $n_{j}\geqslant 2$, where $p_{T, 2nd \, j}> 100$ GeV,  $H_{T,j} >$ 800 GeV and $m_{jj}<$ 45 GeV or $m_{j,j} >$125 GeV
    \item $\Delta R_{Z, j} < 2$ for the nearest jet
  \item $\ETmiss <$ 150 GeV 
  \item $m_{Z,j} > 500$ GeV 
\end{enumerate}

The $t \, \bar{t} + \jets$ background is mostly reduced by cut (1.), $W^{\pm}\, Z^0 + \jets$ by cuts (2.) and (4.), and $Z + \jets$ by (2.), (3.) and (5.). After the full set of cuts, the signal is clearly visible over the background. To estimate the significance for this particular analysis we simply count the number of signal and background events within a window of approximately twice the observed resonance mass. Assuming an integrated luminosity of ${\cal L}= 10\ \text{fb}^{-1}$ we obtain a total background of $B = 84$ events, signal of $S = 217$ events, leading to  $S/ \sqrt{ B + S}= 12.5$ and $S/B= 2.6$. See Appendix \ref{app:singleKK} for more details about the analysis. 

From the observed rate (Fig.~\ref{fig:singKKmass}), one can infer the total cross section for $pp \rightarrow Q_{KK}(Zj) + \jet$ -- a lower bound on the total $Q_{KK} + \jet$ production cross section. While this alone is not enough to pinpoint the individual $\lambda_{l, h^2}$ or $ \lambda_{l^2, h}$, a large value for their product is a strong indication of light fermion compositeness. By studying the correlation among many channels or by going to higher precision machines, perhaps the individual $\lambda_{l^2, h} (\lambda_{l, h^2})$ can be discerned.

For lighter resonances the background is larger and  the signal decay products are less energetic. However the signal cross section is enhanced, both by phase space and because the KK-fermion - KK gluon - zero-mode fermion coupling is higher.


\subsection{Double production fermion KK: the $4\,\ell + 2\ \jet$ channel}
 
KK fermion resonances can also be produced in pairs, both $pp \rightarrow Q_{KK} \, Q_{KK}$ and $pp \rightarrow Q_{KK} \, \bar{Q}_{KK}$ with a large cross section. Unlike single production, pairs of KK quarks can be produced from initial state gluons, so in principle this channel is sensitive to the third generation KK states: $Q_{KK} \,  =  \, U_{KK}\, , \, D_{KK}\, , \, S_{KK}\, , \, C_{KK}\ , \, B_{KK}\ , \, T_{KK}$ and $X$\@. Being lighter, the third generation $Q_{KK}$ do make up a significant fraction of the $s$-channel contribution -- roughly a third ($5\ \pb/15\ \pb$) for the point we consider. However, the total pair production cross section is dominated the $t$-channel piece. Since the $t$-channel diagram is sensitive to the quark distribution functions, the net result is a pair production signal which is predominantly ($\sim 70\%$) first generation KK quarks (and anti-quarks).  Given this bias, in the following we approximate the complete pair production signal with the pair production of $U_{KK}, D_{KK}, \bar U_{KK}$ and $\bar D_{KK}$ alone.

\begin{figure}[h!]
\centering
\begin{minipage}[c]{0.45\textwidth}
\centering
	 \includegraphics[width=2.5in, height=1.3in]{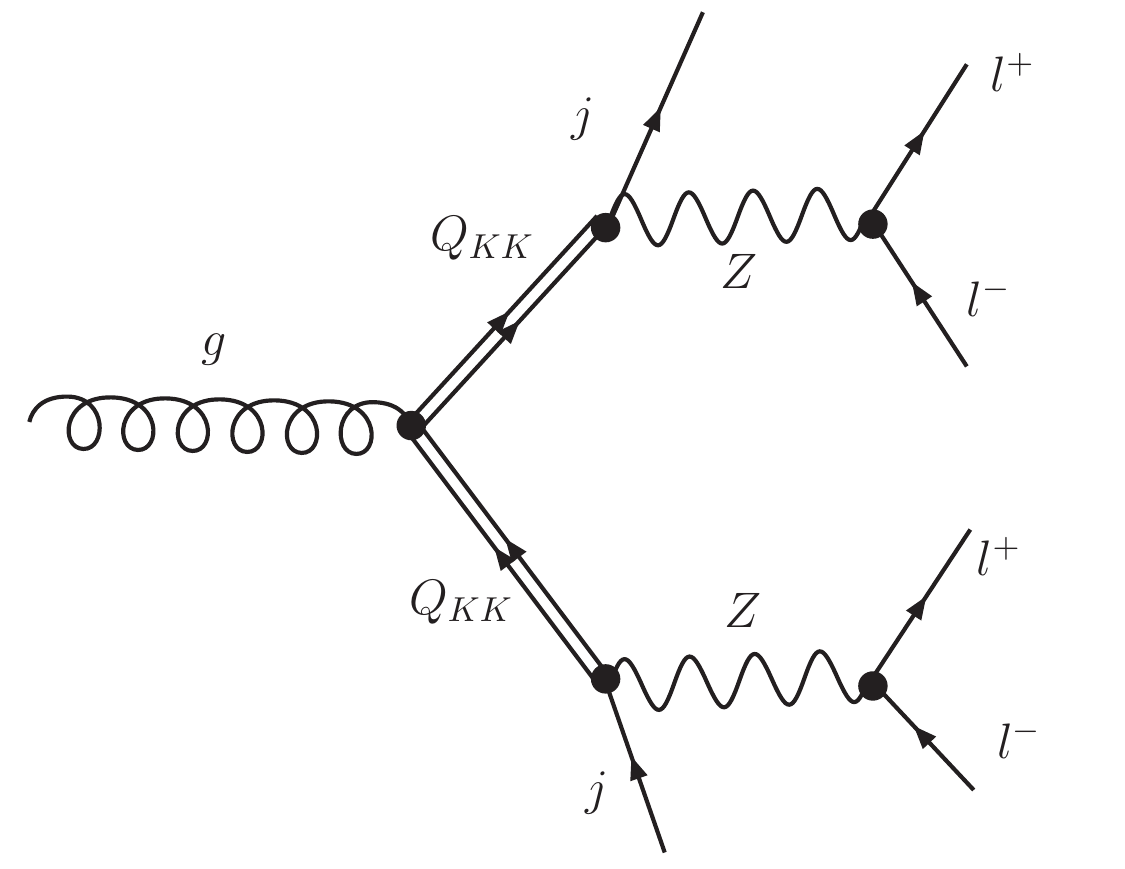}
  	\includegraphics[width=2.in, height=1.1in]{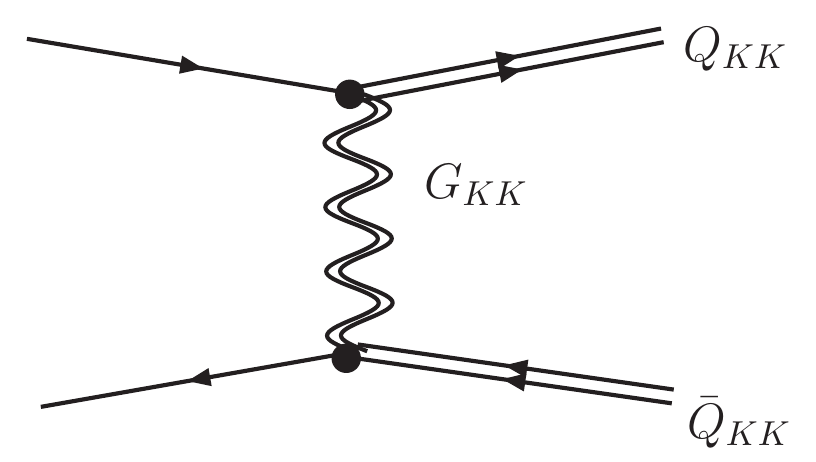}
  	\caption{Two competing contributions to pair production of KK fermions}
	\label{fig:pairdiag}
\end{minipage}
\hspace{0.1in}
\begin{minipage}[c]{0.45\textwidth}
\centering
	 \includegraphics[width=2.5in, height=2.4in]{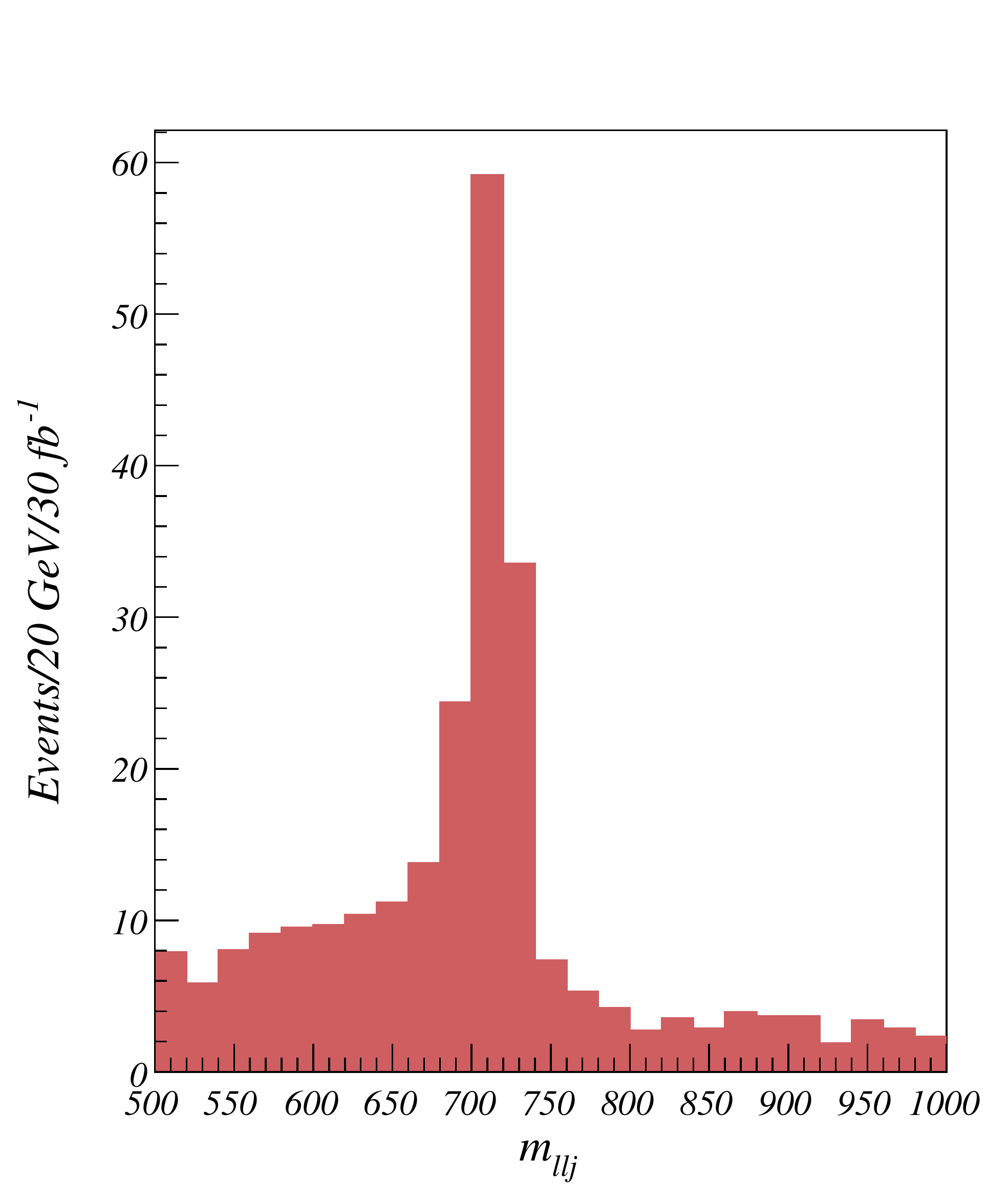}
 	\caption{ Pair production in the 4 $\ell$ + 2 $j$ channel}
	 \label{pairresult}
\end{minipage}
 \end{figure}

With this simplification, the cleanest signature occurs when both KK quarks decay to $Z(\ell^+\ell^-) + \text{light jet}$. The result is a clean signature with 4 leptons and 2 hard jets. Processes with this final state are depicted in Fig.~\ref{fig:pairdiag}.  
The signal, cleaner than the single production, gives a measurement of the KK-fermion mass but has the disadvantage that to obtain information on the size of $\lambda_{l,h^2}$, one has to disentagle the universal gluon contribution in Fig.~\ref{fig:pairdiag}.   
More specifically, the $Q_{KK}$ production in the topmost Feynman diagram proceeds via $s$-channel gluon, whose coupling to the $Q_{KK}$ pairs is fixed by gauge invariance. Knowing this coupling, the $Q_{KK}$ mass and branching ratios, one can theoretically solve for the $t$-channel $G_{KK}$ contribution to the cross section and thereby determine the coupling $\lambda_{l,h^2}$.

We select the final state of interest by imposing the following cuts:
\begin{enumerate}
  \item $n_{\ell}=4$, 2 by 2 same-flavor, opposite sign leptons and $m_{\ell \ell} = m_Z \pm 20$ GeV
  \item $n_{j}\geqslant 2$  with $p_{T,j}> 100$ GeV  
\item $m_{\ell \, \ell \, j}\ \in$[670,750] GeV
\end{enumerate}

The KK fermion is heavy and its decay products tend to be near each other in $\Delta$R parameter space. We used this information to reduce combinatorics: we paired a lepton with the nearest opposite-sign lepton using a $\Delta$R$_{\ell \tilde{\ell}}$ selection.  We followed the same procedure to pair the lepton-lepton pair with the nearest jet and reconstruct the KK fermion mass. In Fig.~\ref{pairresult} we plot one of the $\ell\ell j$ combination reconstruction. 

While there are sources of true SM backgrounds for events with 4 leptons+ 4 jets (such as $ZZ/\gamma\gamma/Z\gamma + 4\ \jets$), their production cross sections are so low that they are rendered completely unimportant. Instead, the relevant backgrounds come from large SM processes, such as $W$+jets, $Z$+jets, and QCD, combined with some of the leptons coming from jet-fakes. Of these fake backgrounds, the largest could potentially be QCD with all four leptons coming from fakes~\cite{DeSimone:2009ws}. Estimating the fake rate with a conservative ( though $p_T$ and $\eta$ independent)  value of $10^{-4}$~\cite{unknown:1999fr, unknown:2006fr}, we can see the requirement of four fake leptons cuts  the QCD rate,  initially $\sim \,10^8\ \pb$, down to well below the $\fb$ level. Similar fake rates applied to backgrounds like $W^{\pm}$+jets or $W^{\pm}Z^0$+jets lead to cross sections significantly below the signal.

Another potential source of background is fully leptonic $\bar t t$, where semileptonic decays of the $b$ quarks can occasionally yield isolated muons. The $t \bar{t}$+jets background has an initial cross section of about $1\ \text{nb}$ and we estimated the $b$-jets produce an isolated lepton in about $5 \times10^{-3}$. Combining the probability for both $b$ quarks to yield isolated leptons with the leptonic branching ratio of the $W$'s, we estimate a $4\,\ell + \text{jets} $ cross section of about $10^{-4}\, \fb$ after imposing cuts (1.) and (2.). 
 
In summary, the large mass and the high lepton multiplicity of the $Q_{KK}$  signal makes it essentially background-free. Cutting around the mass peak ($670-750\ \gev$ range), the signal contains 90 events for 10 fb$^{-1}$ of data, therefore discovery is below $10\ \fb^{-1}$ of data.
   
    
\subsection{Neutral resonances: multiple tiers and nearby states}

 While the large KK gluon-SM fermion-KK fermion coupling is an interesting and useful feature, the strongest indicator of the mass matched scenario is the mass-degeneracy itself. Having illustrated the best channels for $U_{KK}, D_{KK}$ discovery and mass determination in the previous sections, we now focus on the cleanest extraction of the $Z_{KK}$ mass -- the dilepton channel.

%

 Dilepton $Z'-$like resonances have been studied extensively in a number of different scenarios \cite{Cvetic:1997wu,Langacker:2000ju,Chivukula:2002ry, Rizzo:2006nw,Feldman:2006wb,Cata:2009iy}, however they have been overlooked  as a discovery possibility in mass-matched models\footnote{Some preliminary work on this channel was presented in a talk given by Giacomo Cacciapaglia and Guido Marandella at the Budapest meeting.}. In Ref.~\cite{hep-ph/0412278}, mass-matching was assumed to suppress the Drell-Yan signal to the point that vector-boson-fusion and associated production of KK gauge bosons were the preferred discovery modes. In 4D discrete models of mass-matching, so-called `ideally delocalized' models~\cite{Chivukula:2005bn, Chivukula:2006cg, He:2007ge}, the SM-fermion - neutral KK gauge boson coupling is more strongly suppressed than in the 5D models, so the Drell-Yan production of neutral resonances is essentially zero. Drell-Yan neutral resonance production in Higgsless-style models has been investigated in \cite{Accomando:2008dm, Accomando:2008jh,Belyaev:2008yj}, however the mass-matched scenario has several distinct features which we point out here.

For the same signal point as above, we generate the signal for $pp \rightarrow Z_{KK,i} \rightarrow \ell^- \ell^+$, where the $i$ indicates we sum over all kinematically allowed neutral KK states. As usual, $\ell = e, \mu$.  After applying minimum cut of $500\ \text{GeV}$ on the invariant mass of two same-flavor, opposite-sign leptons, the only significant Standard Model background is $pp \rightarrow Z^0/\gamma^* \rightarrow \ell^+ \ell^-$. The large invariant mass cut, along with $p_T$ cuts for the two leptons, $p_T > 200\ \text{GeV}, |\eta| < 2.5$ suppress the background to the point that the signal is easily visible. The invariant mass of the lepton pair for the signal and background are shown below in Fig.~\ref{fig:zll_all}.

\begin{figure}[h!]
\centering
\begin{minipage}[c]{0.3\textwidth}
	\centering
	\includegraphics[width=2.0in, height = 2.5 in]{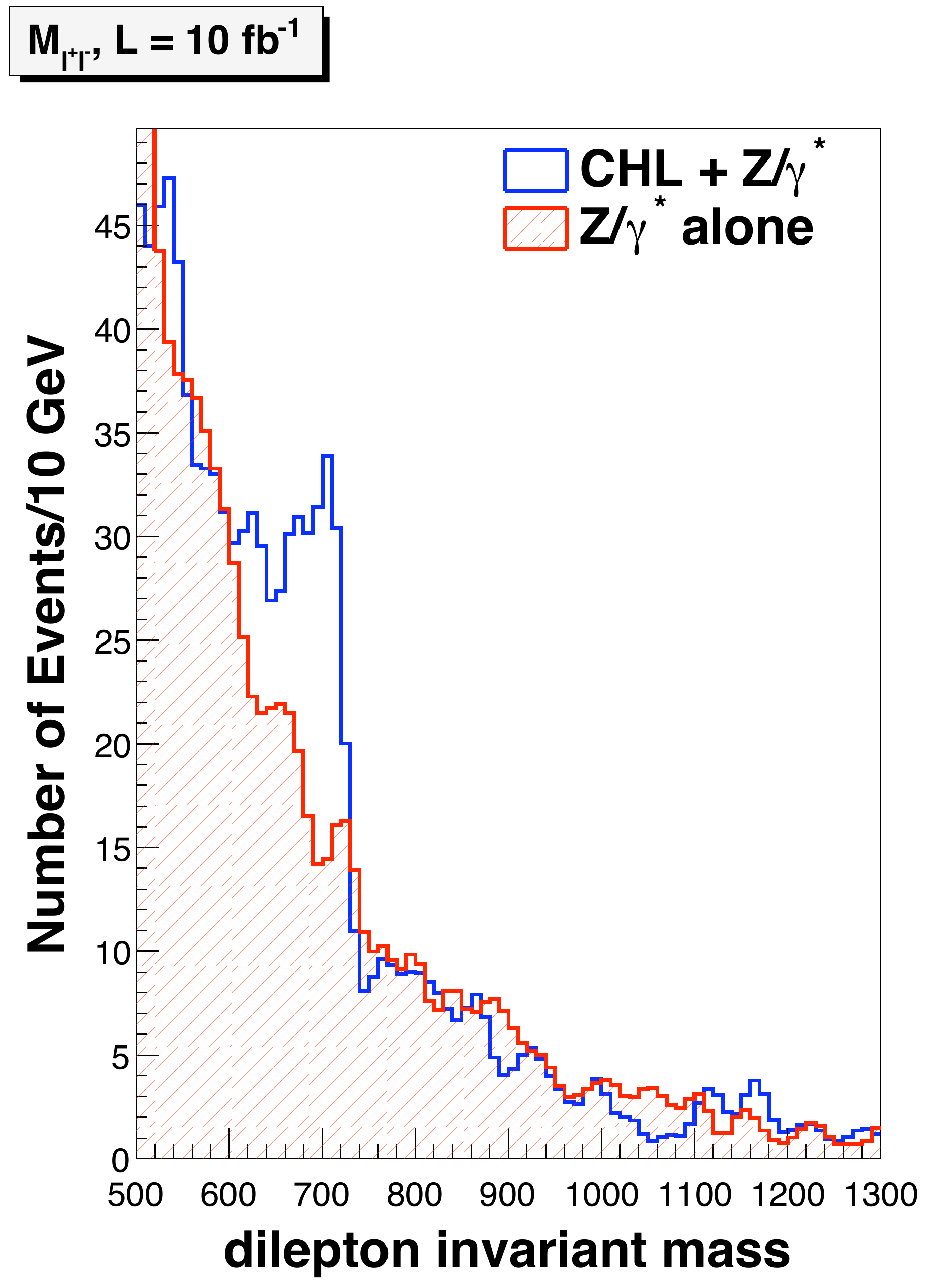}
\end{minipage}
\hspace{0.4cm}
\begin{minipage}[c]{0.3\textwidth}
	\centering
	\includegraphics[width=2.0in, height = 2.5 in]{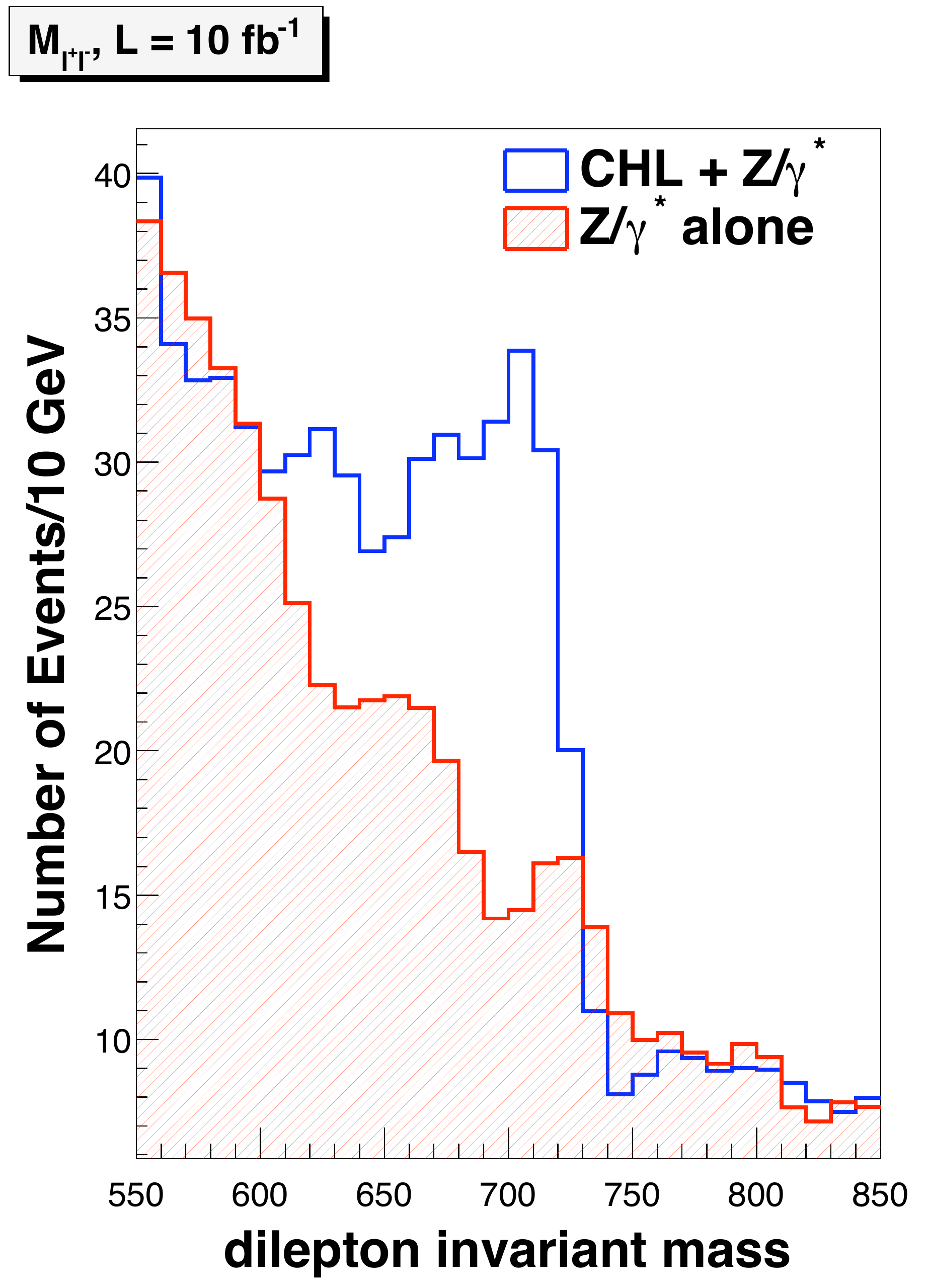}
\end{minipage}
\hspace{0.4cm}
\begin{minipage}[c]{0.3\textwidth}
	\centering
	\includegraphics[width=2.0in, height = 2.5 in]{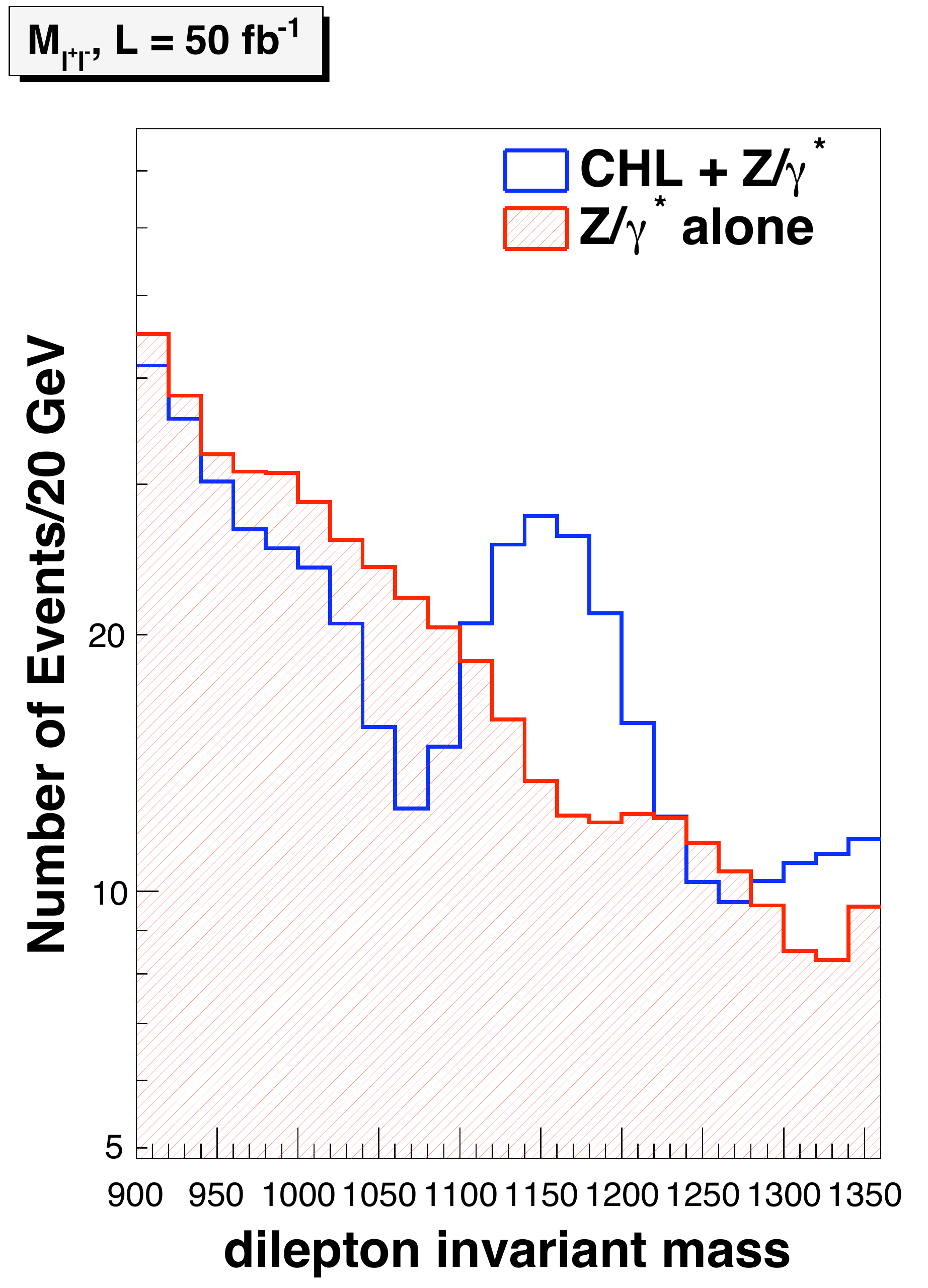}
\end{minipage}
\caption{Neutral resonance reconstruction in the dilepton channel. The solid curve is the signal and background, generated together to maintain any interference between them, while the shaded region is the background alone. A breakdown of the analysis is given in Appendix~(\ref{app:neutapp}). Left: overview in the 500-1300 GeV range. Center: close-up to the two lightest $Z_{KK}$. Right: close-up into the heaviest resonance. Notice the vertical scale in the rightmost plot is logarithmic and the luminosity is five times higher than in the other two plots.}
\label{fig:zll_all}
\end{figure}
There are several features of Fig.~\ref{fig:zll_all} which are different from other $Z'$ searches. First, although the signal is clearly visible at low luminosity, the coupling $g_{ff Z_{KK}}$ is substantially smaller than the SM coupling; typically $g_{ff Z_{KK}} \lesssim 0.2\times g_{f f Z}$. Verifying this small, but nonzero coupling will be necessary to distinguish between mass-matched Higgsless models and other models with a $Z'$.  Another important aspect of the neutral gauge boson resonances in the mass-matched scenario is that they can decay to SM $W^{\pm}$ bosons -- something the $Z'$ from a new $U(1)$ group may not do. Unfortunately the cleanest channel in $pp \rightarrow Z_{KK,i} \rightarrow W^+W^-$ is the fully leptonic channel, where the presence of two sources of missing energy (two neutrinos) makes it impossible to reconstruct the invariant mass. 

The two other distinct features of Fig.~\ref{fig:zll_all} are more difficult to see; they will require, at best, more luminosity, and at worst, a linear collider. These features are the presence of {\em two} nearly degenerate resonances at $\sim 700\ \text{GeV}$, and the sign of the second tier of resonances -- the third $Z_{KK}$ at $\sim 1100\ \text{GeV}$. The nearly degenerate resonances are the KK partner of the Z boson and the KK partner of the photon\cite{Piai:2009da}. For the point we have shown above their separation is $\sim 15\ \text{GeV}$. At a hadron collider, and assuming the EM resolution encoded by PGS, such nearby resonances will be difficult to spot. 

In the figures above, the low-mass peak is dominated by the more massive $Z_{KK}$, with the effects of the lighter resonance hiding as a shoulder. One effect of the nearby resonances which has been completely left out of this plot is interference between the two resonances. Because these two resonances share the same final states, at loop-level the propagator matrix will develop non-zero off-diagonal terms. The effects of these off-diagonal terms has been investigated analytically in Ref.~\cite{Cacciapaglia:2009ic}~\footnote{Current Monte Carlo programs, being tree level tools, do not include incorporate interference effects for generic resonances. For the specific case of the multiple Higgses of the MSSM, however, programs do exist~\cite{Heinemeyer:1998yj} }, and can lead to substantial changes in signal rate and shape, depending on the mass of the resonances. For the particular point above, these effects are estimated to be at the few percent level, but for higher masses the effect is ${\cal O}(1)$, not only in the total cross section but in the shape of the peaks~\cite{Cacciapaglia:2009ic}.

The presence of a multiple tiers of resonance is a strong indicator of physics with an extra-dimensional (or large-$N_C$) description at work. As they are more massive, the higher tier resonances have more open decay possibilities and thus they tend to be wide; for the point shown above the tree-level width for the third neutral resonance $Z_{KK,3}$ is $\sim 90\ \text{GeV}$. Despite their large mass, the second-tier resonances are actually more strongly coupled to the zero-mode fermions. A  larger coupling means the LHC reach may extend farther than naively expected. From the rightmost plot in Fig.~\ref{fig:zll_all} it appears that these higher-tier resonances will be visible at the LHC for a wide range of mass-matched Higgsless parameter space, however discovery requires substantially higher luminosity. Also, as detection of the $Z_{KK,3}$ involves isolating a rather small,  wide peak over a swiftly falling SM background, our analysis becomes quite sensitive to the details of detector response (EM resolution, difference between electrons and muons, etc.). Therefore a dedicated study using more vetted tools is needed to accurately determine the LHC reach for the $Z_{KK,3}$ and other higher-tier KK states.


\subsection{Charged resonances:  $3\,\ell + \ETmiss$}
\label{sec:chargeres}

In the previous section, we discussed the determination of the $Z_{KK}$ resonance masses -- a necessary ingredient to identify mass-matching. For completeness, in this section we present the discovery of $W_{KK}$ excitations in the fully leptonic $W^{\pm} \, Z^0$ channel.

 The $Z_{KK}$ and $W_{KK}$ masses differ by few GeV because they form an approximate $SU(2)$ triplet.
This $SU(2)$ symmetry is preserved by the IR brane, and only broken by UV boundary conditions. KK resonances are rather insensitive to UV breaking, hence the small splitting\cite{Hirn:2007bb}.  The phenomenology of a $W_{KK}$ in this channel has been considered before, see for example Refs.\cite{Hirn:2007we, Ohl:2008ri, Cata:2009iy, Alves:2009aa}. Its phenomenology is also very similar to the $\rho_T$ of Low-Scale Technicolor~\cite{Eichten:1996dx, Eichten:1997yq} or the heavy $W'$ is deconstructed Higgsless models~\cite{He:2007ge, Alves:2008up}.

The important backgrounds for this process are, $W^{\pm} \, Z^0$ $\rightarrow$ 3$\ell$ + $\nu$ (irreducible), $Z^0\, Z^0 \rightarrow 4 \ell$ where a lepton is missed, $Z^0+ b \bar{b}$ and $t \bar{t}$. We selected the events with the following cuts:
\begin{enumerate}
  \item $n_{\ell}=3$ and look for two opposite-sign same-flavor leptons with invariant mass closest to the $Z^0$.
  \item Reconstruct the neutrino momentum by pairing the remaining lepton with the missing energy to reconstruct the $W$. There is a twofold ambiguity which we resolve by asking for the neutrino solution most collinear with the unpaired lepton.  
  \item Kinematic cuts: $p_{T,W}> $ 200 GeV, $p_{T,Z}>$ 250 GeV and $H_{T,j}< 125\ \gev$.
\end{enumerate}
The reconstruction of the three leptons and missing energy combination is shown in Fig.~\ref{charged} below. The peak for $W_{KK}$  is clearly visible -- after applying an invariant mass cut of 590 GeV $< m_{W\, Z} <$ 820 GeV, there are 84 signal events, 17 background events for a luminosity of 10 fb$^{-1}$. See Appendix \ref{chargedapp} for more details on the background and signal efficiencies to the cuts.
 \begin{figure}[h!]
\begin{center}
   \includegraphics[width=2.2in, height=3.0in, angle=90]{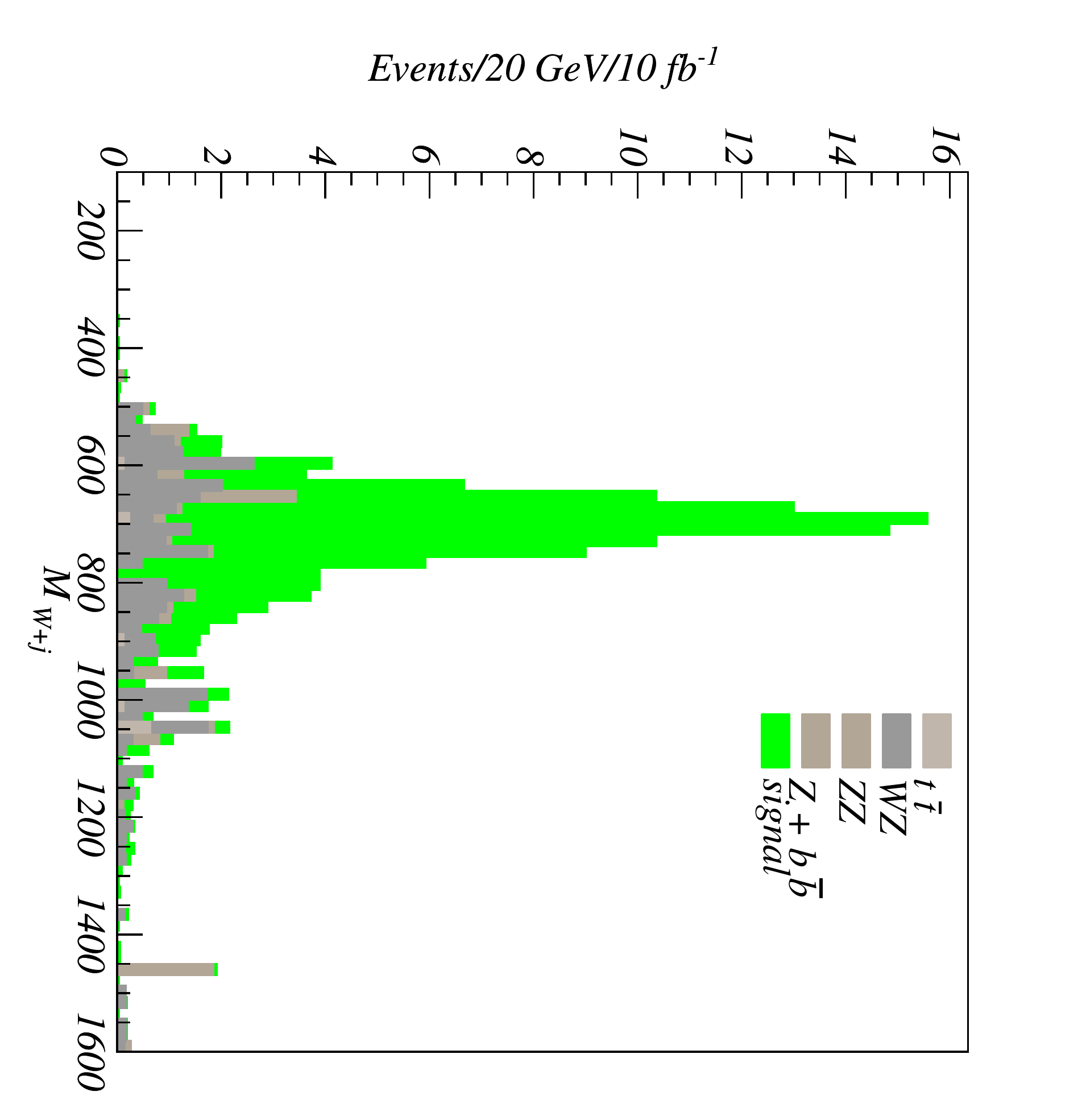}
 \caption{Charged resonance reconstruction in the 3 $\ell$ + $\ETmiss$ channel.}
 \label{charged}
 \end{center}
 \end{figure}

\section{Conclusions}

Higgsless models rely on light spin one resonances instead of the Higgs boson in order to perturbatively unitarize $W_L W_L$ scattering.
As a drawback,  light resonances participating in EWSB yield appreciable contributions to the $S$ parameter\cite{Peskin:1990zt,Barbieri:2004qk}.
No known symmetry protects a model from those dangerous contributions, but the dynamics of some models might lead to relative cancellations of competing contributions to the $S$ parameter.
Such scenarios can be best described in an extra-dimensional framework.

The Cured Higgsless model is an example of such a proposal. Assuming some tuning, contributions to $S$ from different sectors in the theory -- fermionic and bosonic-- balance each other in $S$\@. No symmetry protects these cancellations, but if they are produced by the dynamics of a strongly-interacting scenario, the hope is that the extra-dimensional description can predict the concomittant phenomenology. Indeed, we find that the cancellation is realized as a mass-matching between fermionic and bosonic KK resonances. To achieve this, it is necessary that the SM light quarks and leptons be partially composite.

Thus, measuring similar masses for KK-quarks and KK-gauge bosons is a hallmark of the Cured Higgsless scenario. Also, another striking consequence of the compositeness of SM fermions is a sizable trilinear couplings among KK  gluonic resonances and light quarks.

We studied the reach of a benchmark model with resonances in the $700\ \gev$ range. In that model, fermionic KK resonances decay predominantly to gauge bosons and quarks. We focused on the cleanest signals, where the KK fermion decays to leptonic $Z$'s. KK fermions can be singly or pair produced, leading to 2 $\ell$+ 2 $j$  and 4 $\ell$+ 2 $j$  final states, respectively. Bosonic KK resonances form an approximate $SU(2)$ triplet, where neutral and charged resonances have similar masses. For the neutral resonance we choose the dilepton final state, whereas we opted for the leptonic $W^{\pm}\, Z^0$ for the charged resonance. In both the charged and neutral cases we considered the $s$-channel production, which had been overlooked for the Cured Higgsless scenario but it actually dominates for light resonances.

All of these channels present a discovery within the 10 fb$^{-1}$ of ($\sqrt s = 14\ \tev$) data, once backgrounds are taken into consideration. Comparing the extracted mass of KK fermions and KK bosons, one can check the mass-matching premise. In addition, measuring a significant single production of the KK fermion is a clear indication of compositeness in the SM fermions, a key piece in the Cured Higgsless proposal.

\label{sec:conclude}
\section*{Acknowledgements}

We thank Giacomo Cacciapaglia for collaboration in the early stages of this work and useful comments. We also like to thank him, Aldo de Andrea and Stefania de Curtis for sharing their results on the interference of nearby resonances\cite{Cacciapaglia:2009ic}. We would like to thank Csaba Csaki and Sekhar Chivukula for useful comments and encouragement, and Piyali Banerjee for discussions on the dilepton channel. The authors also enjoyed the environment in the Aspen Center  of Physics, where this work was initiated.
AM is supported by the DoE under grant No. DE-FG02-92ER-40704, while VS under grant No. DE-FG02-91ER40676 and NSERC funding from the government of Canada.

%
%
\begin{appendix}
\renewcommand{\theequation}{A-\arabic{equation}}
\setcounter{equation}{0} 
\renewcommand{\thetable}{A.\arabic{table}}
\setcounter{table}{0}

\section{Appendix: Cured Higgsless Parameters}
\label{CHLPAR}
The KK spectrum which results from the 5D parameters in Equation~\ref{eq:CHLPOINT} is shown in Table \ref{table:KKSPECT}. 

\begin{table}[h!]
\begin{minipage}[c]{0.3\textwidth}
\centering
\begin{tabular}{ |c|c|}\hline
KK Gauge Bosons &  \\ \hline
$G_{KK}$ & $717.5\ \gev$  \\ \hline
$W_{KK,1}$ & $699\ \gev$ \\ \hline
$W_{KK,2}$ & $1105\ \gev$ \\ \hline
$Z_{KK, 1}$ & $694\ \gev$ \\ \hline
$Z_{KK, 2}$ & $717\ \gev$ \\ \hline
$Z_{KK, 3}$ & $1111\ \gev$ \\ \hline
\end{tabular}
\label{table:KKSPECT}
\end{minipage}
\hspace{0.4in}
\begin{minipage}[c]{0.3\textwidth}
\centering
\begin{tabular}{|c|c|}\hline
 KK Leptons &  \\ \hline
 $E_{KK, 1,2}$ & $722\ \gev$  \\ \hline
 $E_{KK, 3}$ & $656\ \gev$  \\ \hline
 $N_{KK, 1,2}$ & $722\ \gev$ \\ \hline
 $N_{KK,3}$ & $657\ \gev$ \\ \hline
\end{tabular}
\label{table:KKSPECT2}
\end{minipage}
\hspace{-0.2in}
\begin{minipage}[c]{0.3\textwidth}
\centering
\begin{tabular}{ |c|c|}\hline
 KK Quark & \\ \hline
$U_{KK}$& $722\ \gev$ \\ \hline
$D_{KK}$ & $722\ \gev$ \\ \hline
$C_{KK}$& $702\ \gev$ \\ \hline
$S_{KK}$& $698\ \gev$\\ \hline
$B_{KK,1}$ & $595\ \gev$\\ \hline
$B_{KK,2}$ & $935\ \gev$\\ \hline
$T_{KK,1}$ & $541\ \gev$ \\ \hline
$T_{KK,2}$ & $870\ \gev$ \\ \hline
$X_{KK,1}$ & $460\ \gev$ \\ \hline
\end{tabular}
\end{minipage}
\caption{First KK masses for the parameter set above. The $X$ fermion is the charge $+\frac{5}{3}$ quark which accompanies the third generation quarks under the charge assignment in~\cite{Agashe:2006at}.}
\label{table:KKSPECT3}
\end{table}
\renewcommand{\thetable}{B.\arabic{table}}
\section{Appendix:  Simulation details}
\subsection*{$2\,\ell + 2\ \text{jet}$ channel}
\label{app:singleKK}

The dominant Standard Model backgrounds to this final state are $Z (\ell^+\ell^-) + \text{jets}$, fully leptonic $t\bar t$, and $W/Z + \text{jets}$. The details of the Monte Carlo simulation of these backgrounds, including the cross section and the derived efficiency to pass the cuts described in the text  are presented below. Unless otherwise stated, all backgrounds were created using ALPGENv13~\cite{Mangano:2002ea} with the default parton distribution functions (CTEQ5L) and renormalization/factorization schemes. The signal  -- the point in table~(\ref{table:KKSPECT3}) -- was generated in MadGraphv4 using a fixed renormalization/factorization scheme of $700\ \gev$. For subprocesses with $+0\ \jets$ through $+2\ \jets$ we used the MLM jet-parton matching scheme built into ALPGEN, while the $+3^+\ \jets$ subprocess is generated inclusively.

The total \# events for each process, although only used in the normalization of histograms, has been chosen to roughly correspond with the number of events in $\mathcal L = 10\ \fb^{-1}$. The quoted $\epsilon$ is the efficiency after the cuts described in Section ~(\ref{ssec:singleKK}), combined with the object identification/reconstruction efficiency of PGS\@.  The last column indicates the number of signal and background events within a mass window of $m_{Z,j} \in [660, 780]\ \gev$ (numbers are given only for those backgrounds which were nonzero before the mass window). The last column was used to calculate the significance quoted in the text.
\begin{centering}
\begin{table}[h!]
\centering 
\begin{tabular}{ |c|c|c|c|c|c| }\hline
Process & $\sigma$(pb) & total \# events & $\epsilon $ & total \# :  & \# mass peak \\
& & & & $(10\ \fb^{-1}) \cdot \sigma \cdot \epsilon$ & \\
\hline \hline & & &  & &\\ 
$Z(\ell^+\ell^-) + 0\ \text{jets}$ & $3946\ \text{pb}$ & $2.76\times 10^6$ & 0.0 & 0 & \\
$Z(\ell^+\ell^-) + 1\ \text{jets}$ & $1737\ \text{pb}$ & $1.27\times 10^7$ & $1.8\times 10^{-6}$ & 32 & 1 \\
$Z(\ell^+\ell^-) + 2\ \text{jets}$ & $967\ \text{pb}$ & $6.43\times 10^6$ & $8.4\times 10^{-5}$ & 244 & 42\\
$Z(\ell^+\ell^-) + 3^+\ \text{jets}$ & $291\ \text{pb}$ & $2.93\times 10^6$ & $3.0\times 10^{-5}$ & 290 & 38\\
& & & & &\\ 
$t\bar t + 0\ \text{jets}$ & $443\ \pb$ & $2.25\times 10^7$& 0& 0 &\\
$t\bar t + 1\ \text{jets}$ & $750\ \pb$& $7.09\times 10^6$& 0 & 0 &\\
$t\bar t + 2\ \text{jets}$ & $778\ \pb$& $4.09 \times 10^6$& 0 & 0 &\\
$t\bar t + 3^+\ \text{jets}$ & $254\ \pb$& $3.54\times 10^6$& $5.65\times 10^{-7} $& $\le 1$ & 0\\
& & & & &\\
$WZ + 0\ \text{jets}$ & $28.1\ \pb$& $1.38\times 10^6$& 0 & 0 & \\
$WZ + 1\ \text{jets}$ & $32.2\ \pb$& $1.45\times 10^6$& $2.75\times 10^{-6}$ & 1 & 0\\
$WZ + 2\ \text{jets}$ & $28.9\ \pb$& $3.44\times 10^5$& $1.8\times 10^{-5}$ & 8 & 2\\
$WZ + 3^+\ \text{jets}$ & $11\ \pb$ & $1.22\times 10^5$& $8.2\times 10^{-5}$ & 15 & 1\\
\hline
Total Background & & & & 582 & 84\\ \hline
Signal & $0.322\ \pb$ & $10^4$ & $0.091$ & 293 & 217\\ \hline
\end{tabular}
\caption{Signal and backgrouds to the final state $2\ell + 2\ \text{jet}$. A final number of events $\le 1$ means at least one event passed the cuts, but after proper normalization the number of events in $\mathcal L = 10\ \fb^{-1}$ was less than one. For all processes, the following parton-level cuts have been imposed: for jets: $p_T > 15\ \text{GeV}, |\eta| < 4.0, \Delta R_{jj} > 0.4$, and for leptons: $p_T > 10\ \text{GeV}, |\eta| < 4.0, \Delta R_{\ell\ell}, \Delta R_{\ell-j} > 0.4$. All backgrounds were generated with ALPGENv13 using default parton distribution sets and factorization/renormalization scales.}
\label{table:sigbr}
\end{table}
\end{centering}
\subsection*{Neutral resonance channel}
\label{app:neutapp}

Here we present the details of the neutral resonance search for the Cured Higgsless point explored in the text.
The dominant SM background in the neutral resonance channel is $Z/\gamma^*$. All other backgrounds are percent level effects, which we neglect here. The spectrum of the signal point is shown in Section~(\ref{CHLPAR}).

In order to maintain any interference between the SM and the neutral resonances, we generated the events for this channel in a slightly different way. First, the signal and background were generated together using MadGraphv4 and imposing the parton-level cuts $M_{\ell^+\ell^-} > 300\ \gev, p_{T, \ell} > 150\ \gev$. The combined distributions were then compared with SM-only distributions, generated using exactly the same cuts and parameters. The cross sections and efficiencies we get from this procedure are shown below in table~(\ref{table:sigbrneu}).

\begin{centering}
\begin{table}[!ht]
\centering 
\begin{tabular}{ |c|c|c|c|c|c| }\hline
Process & $\sigma$(pb) & total \# events & $\epsilon $ & total \# :  & \# mass peak \\
& & & & $(10\ \fb^{-1}) \cdot \sigma \cdot \epsilon$ & \\
\hline \hline & & &  & &\\ 
$\gamma^*/Z(\ell^+ \ell^-)$ & $0.59 \ \pb$ & $2\times 10^4$ & $0.184$ & $941$ & 375 \\ \hline
CHL + $\gamma^*/Z(\ell^+ \ell^-)$ & $0.61 \ \pb$ & $2\times 10^4$ & $0.202$ & $1049$ & 493 \\ \hline
Difference & & & & 106 & 118 \\ \hline
\end{tabular}
\caption{Signal and background in the neutral resonance channel $\ell^+\ell^-$. Other than the parton-level dilepton invariant mass cut and lepton $p_T$ cuts, all other simulation inputs are exactly as in Table (\ref{table:sigbr}), except MadGraphv4 has been used instead of ALPGEN.}
\label{table:sigbrneu}
\end{table}
\end{centering}

\subsection*{Charged resonance channel}
\label{chargedapp}

Finally, we give the details of the charged KK gauge boson search. The backgrounds we considered are $t\bar t + \text{jets},\ W^{\pm}Z^0 + \text{jets},\ Z^0Z^0 + \text{jets}$ and $Z^0 + \bar b b$. As before, the second to last column indicates the number of events after the cuts presented in section (\ref{sec:chargeres}), while the final column contains the number of events with an additional mass window cut $m_{W,Z} \in [590, 820]\ \gev$.

\begin{centering}
\begin{table}[h!]
\centering 
\begin{tabular}{ |c|c|c|c|c|c| }\hline
Process & $\sigma$(pb) & total \# events & $\epsilon $ & total \# :  & \# mass peak \\
& & & & $(10\ \fb^{-1}) \cdot \sigma \cdot \epsilon$ & \\
\hline \hline & & &  & &\\  
$t\bar t + 0\ \text{jets}$ & $443\ \pb$ & $2.25\times 10^7$&  $2.22\times 10^{-7}$& $\le 1$ & 0 \\
$t\bar t + 1\ \text{jets}$ & $750\ \pb$& $7.09\times 10^6$& $1.41\times 10^{-7}$ & $\le 1$ & 0\\
$t\bar t + 2\ \text{jets}$ & $778\ \pb$& $4.09 \times 10^6$& 0 & 0 & \\
$t\bar t + 3^+\ \text{jets}$ & $254\ \pb$& $3.54\times 10^6$&  0 & 0&  \\
& & & & \\
$WZ + 0\ \text{jets}$ & $28.1\ \pb$& $1.38\times 10^6$& $4.61\times 10^{-5}$ & 10 & 4\\
$WZ + 1\ \text{jets}$ & $32.2\ \pb$& $1.45\times 10^6$& $5.63\times 10^{-5}$ & 13 & 6\\
$WZ + 2\ \text{jets}$ & $28.9\ \pb$& $3.44\times 10^5$& $4.06\times 10^{-5}$ & 9 & 4\\
$WZ + 3^+\ \text{jets}$ & $11\ \pb$ & $1.22\times 10^5$&  $8.19\times 10^{-6}$& $\le 1$ & 0\\
& & & & \\
$ZZ + 0\ \text{jets}$ & $10.9\ \pb$& $1.70\times 10^6$& $1.11\times 10^{-5}$ & 2 & $\le 1$\\
$ZZ + 1\ \text{jets}$ & $8.7\ \pb$& $3.09\times 10^5$& $1.29\times 10^{-5}$ & 2 & $\le 1$\\
$ZZ + 2\ \text{jets}$ & $5.6\ \pb$& $5.53\times 10^4$& $3.62\times 10^{-5}$ & 4 & 2\\
$ZZ + 3^+\ \text{jets}$ & $1.6\ \pb$ & $1.51\times 10^4$& 0 & 0 & \\
& & & & \\
$Z(\bar{\nu}\nu) + \bar b b$ & $24.7\ \pb$ & $6.30\times 10^5$ & 0 &  0 & \\ 
\hline
Total Background & & & & 41& 18\\ \hline
Signal & $0.029 \pb$& $10^4$ & 0.34 & 100 & 85\\ \hline
\end{tabular}
\caption{Backgrounds for the charged KK gauge boson search. Parton level cuts and simulation inputs are the same as in Table~(\ref{table:sigbr}). A final number of events $\le 1$ means at least one event passed the cuts, but after proper normalization the number of events in $\mathcal L = 10\ \fb^{-1}$ was less than one. }
\label{table:sigbr2}
\end{table}
\end{centering}
\end{appendix}
\vfil\eject

\bibliographystyle{utcaps}
\bibliography{massmatch}

\providecommand{\href}[2]{#2}\begingroup\raggedright\begin{thebibliography}{10}

\bibitem{Susskind:1978ms}
L.~Susskind, ``Dynamics of Spontaneous Symmetry Breaking in the Weinberg-Salam
  Theory,'' {\em Phys. Rev.} {\bf D20} (1979) 2619--2625.

\bibitem{Weinberg:1979bn}
S.~Weinberg, ``Implications of Dynamical Symmetry Breaking: an addendum,'' {\em
  Phys. Rev.} {\bf D19} (1979) 1277--1280.

\bibitem{Peskin:1990zt}
M.~E. Peskin and T.~Takeuchi, ``A new constraint on a strongly interacting
  Higgs sector,'' {\em Phys. Rev. Lett.} {\bf 65} (1990) 964--967.

\bibitem{Golden:1990ig}
M.~Golden and L.~Randall, ``Radiative corrections to electroweak parameters in
  technicolor theories,'' {\em Nucl. Phys.} {\bf B361} (1991) 3--23.

\bibitem{Holdom:1990tc}
B.~Holdom and J.~Terning, ``Large corrections to electroweak parameters in
  technicolor theories,'' {\em Phys. Lett.} {\bf B247} (1990) 88--92.

\bibitem{Holdom:1981rm}
B.~Holdom, ``Raising the sideways scale,'' {\em Phys. Rev.} {\bf D24} (1981)
  1441.

\bibitem{Yamawaki:1986zg}
K.~Yamawaki, M.~Bando, and K.-i. Matumoto, ``Scale invariant technicolor model
  and a technidilaton,'' {\em Phys. Rev. Lett.} {\bf 56} (1986) 1335.

\bibitem{Appelquist:1986an}
T.~W. Appelquist, D.~Karabali, and L.~C.~R. Wijewardhana, ``Chiral Hierarchies
  and the Flavor Changing Neutral Current Problem in Technicolor,'' {\em Phys.
  Rev. Lett.} {\bf 57} (1986) 957.

\bibitem{hep-ph/9206225}
R.~Sundrum and S.~D.~H. Hsu, ``Walking technicolor and electroweak radiative
  corrections,'' {\em Nucl. Phys.} {\bf B391} (1993) 127--146,
  \href{http://xxx.lanl.gov/abs/hep-ph/9206225}{ hep-ph/9206225}.

\bibitem{Lane:1994pg}
K.~D. Lane, ``{Technicolor and precision tests of the electroweak
  interactions},'' \href{http://xxx.lanl.gov/abs/hep-ph/9409304}{
  hep-ph/9409304}.

\bibitem{Appelquist:1998xf}
T.~Appelquist and F.~Sannino, ``The physical spectrum of conformal SU(N) gauge
  theories,'' {\em Phys. Rev.} {\bf D59} (1999) 067702,
  \href{http://xxx.lanl.gov/abs/hep-ph/9806409}{ hep-ph/9806409}.

\bibitem{Kurachi:2006mu}
M.~Kurachi and R.~Shrock, ``{Behavior of the S parameter in the crossover
  region between walking and QCD-like regimes of an SU(N) gauge theory},'' {\em
  Phys. Rev.} {\bf D74} (2006) 056003,
  \href{http://xxx.lanl.gov/abs/hep-ph/0607231}{ hep-ph/0607231}.

\bibitem{Sikivie:1980hm}
P.~Sikivie, L.~Susskind, M.~B. Voloshin, and V.~I. Zakharov, ``{Isospin
  Breaking in Technicolor Models},'' {\em Nucl. Phys.} {\bf B173} (1980) 189.

\bibitem{Terning:1994sc}
J.~Terning, ``{Chiral technicolor and precision electroweak measurements},''
  {\em Phys. Lett.} {\bf B344} (1995) 279--286,
  \href{http://xxx.lanl.gov/abs/hep-ph/9410233}{ hep-ph/9410233}.

\bibitem{Agashe:2003zs}
K.~Agashe, A.~Delgado, M.~J. May, and R.~Sundrum, ``RS1, custodial isospin and
  precision tests,'' {\em JHEP} {\bf 08} (2003) 050,
  \href{http://xxx.lanl.gov/abs/hep-ph/0308036}{ hep-ph/0308036}.

\bibitem{Randall:1999ee}
L.~Randall and R.~Sundrum, ``{A large mass hierarchy from a small extra
  dimension},'' {\em Phys. Rev. Lett.} {\bf 83} (1999) 3370--3373,
  \href{http://xxx.lanl.gov/abs/hep-ph/9905221}{ hep-ph/9905221}.

\bibitem{Grossman:1999ra}
Y.~Grossman and M.~Neubert, ``{Neutrino masses and mixings in non-factorizable
  geometry},'' {\em Phys. Lett.} {\bf B474} (2000) 361--371,
  \href{http://xxx.lanl.gov/abs/hep-ph/9912408}{ hep-ph/9912408}.

\bibitem{Gherghetta:2000kr}
T.~Gherghetta and A.~Pomarol, ``{A warped supersymmetric standard model},''
  {\em Nucl. Phys.} {\bf B602} (2001) 3--22,
  \href{http://xxx.lanl.gov/abs/hep-ph/0012378}{ hep-ph/0012378}.

\bibitem{Cacciapaglia:2007fw}
G.~Cacciapaglia {\em et.~al.}, ``{A GIM Mechanism from Extra Dimensions},''
  {\em JHEP} {\bf 04} (2008) 006, \href{http://xxx.lanl.gov/abs/0709.1714}{
  0709.1714}.

\bibitem{Hirn:2006nt}
J.~Hirn and V.~Sanz, ``A negative S parameter from holographic technicolor,''
  {\em Phys. Rev. Lett.} {\bf 97} (2006) 121803,
  \href{http://xxx.lanl.gov/abs/hep-ph/0606086}{ hep-ph/0606086}.

\bibitem{Hirn:2006wg}
J.~Hirn and V.~Sanz, ``The fifth dimension as an analogue computer for strong
  interactions at the LHC,'' \href{http://xxx.lanl.gov/abs/hep-ph/0612239}{
  hep-ph/0612239}.

\bibitem{Cacciapaglia:2004rb}
G.~Cacciapaglia, C.~Csaki, C.~Grojean, and J.~Terning, ``Curing the ills of
  Higgsless models: The S parameter and unitarity,'' {\em Phys. Rev.} {\bf D71}
  (2005) 035015, \href{http://xxx.lanl.gov/abs/hep-ph/0409126}{
  hep-ph/0409126}.

\bibitem{SekharChivukula:2008gz}
R.~S. Chivukula and E.~H. Simmons, ``{A Four-site Higgsless Model with
  Wavefunction Mixing},'' {\em Phys. Rev.} {\bf D78} (2008) 077701,
  \href{http://xxx.lanl.gov/abs/0808.2071}{ 0808.2071}.

\bibitem{Eichten:2007sx}
E.~Eichten and K.~Lane, ``{Low-scale technicolor at the Tevatron and LHC},''
  \href{http://xxx.lanl.gov/abs/0706.2339}{ 0706.2339}.

\bibitem{Foadi:2007ue}
R.~Foadi, M.~T. Frandsen, T.~A. Ryttov, and F.~Sannino, ``{Minimal Walking
  Technicolor: Set Up for Collider Physics},'' {\em Phys. Rev.} {\bf D76}
  (2007) 055005, \href{http://xxx.lanl.gov/abs/0706.1696}{ 0706.1696}.

\bibitem{Dietrich:2008up}
D.~D. Dietrich and C.~Kouvaris, ``{Generalised bottom-up holography and walking
  technicolour},'' {\em Phys. Rev.} {\bf D79} (2009) 075004,
  \href{http://xxx.lanl.gov/abs/0809.1324}{ 0809.1324}.

\bibitem{Nunez:2008wi}
C.~Nunez, I.~Papadimitriou, and M.~Piai, ``{Walking Dynamics from String
  Duals},'' \href{http://xxx.lanl.gov/abs/0812.3655}{ 0812.3655}.

\bibitem{Gubser:1998bc}
S.~S. Gubser, I.~R. Klebanov, and A.~M. Polyakov, ``{Gauge theory correlators
  from non-critical string theory},'' {\em Phys. Lett.} {\bf B428} (1998)
  105--114, \href{http://xxx.lanl.gov/abs/hep-th/9802109}{ hep-th/9802109}.

\bibitem{Witten:1998qj}
E.~Witten, ``{Anti-de Sitter space and holography},'' {\em Adv. Theor. Math.
  Phys.} {\bf 2} (1998) 253--291,
  \href{http://xxx.lanl.gov/abs/hep-th/9802150}{ hep-th/9802150}.

\bibitem{Pomarol:2000hp}
A.~Pomarol, ``{Grand Unified Theories without the Desert},'' {\em Phys. Rev.
  Lett.} {\bf 85} (2000) 4004--4007,
  \href{http://xxx.lanl.gov/abs/hep-ph/0005293}{ hep-ph/0005293}.

\bibitem{ArkaniHamed:2000ds}
N.~Arkani-Hamed, M.~Porrati, and L.~Randall, ``{Holography and
  phenomenology},'' {\em JHEP} {\bf 08} (2001) 017,
  \href{http://xxx.lanl.gov/abs/hep-th/0012148}{ hep-th/0012148}.

\bibitem{Csaki:2003dt}
C.~Csaki, C.~Grojean, H.~Murayama, L.~Pilo, and J.~Terning, ``Gauge theories on
  an interval: Unitarity without a Higgs,'' {\em Phys. Rev.} {\bf D69} (2004)
  055006, \href{http://xxx.lanl.gov/abs/hep-ph/0305237}{ hep-ph/0305237}.

\bibitem{Csaki:2003zu}
C.~Csaki, C.~Grojean, L.~Pilo, and J.~Terning, ``Towards a realistic model of
  Higgsless electroweak symmetry breaking,'' {\em Phys. Rev. Lett.} {\bf 92}
  (2004) 101802, \href{http://xxx.lanl.gov/abs/hep-ph/0308038}{
  hep-ph/0308038}.

\bibitem{Nomura:2003du}
Y.~Nomura, ``{Higgsless theory of electroweak symmetry breaking from warped
  space},'' {\em JHEP} {\bf 11} (2003) 050,
  \href{http://xxx.lanl.gov/abs/hep-ph/0309189}{ hep-ph/0309189}.

\bibitem{Hirn:2005nr}
J.~Hirn and V.~Sanz, ``{Interpolating between low and high energy QCD via a 5D
  Yang-Mills model},'' {\em JHEP} {\bf 12} (2005) 030,
  \href{http://xxx.lanl.gov/abs/hep-ph/0507049}{ hep-ph/0507049}.

\bibitem{Hirn:2005vk}
J.~Hirn, N.~Rius, and V.~Sanz, ``{Geometric approach to condensates in
  holographic QCD},'' {\em Phys. Rev.} {\bf D73} (2006) 085005,
  \href{http://xxx.lanl.gov/abs/hep-ph/0512240}{ hep-ph/0512240}.

\bibitem{Policastro:2002se}
G.~Policastro, D.~T. Son, and A.~O. Starinets, ``{From AdS/CFT correspondence
  to hydrodynamics},'' {\em JHEP} {\bf 09} (2002) 043,
  \href{http://xxx.lanl.gov/abs/hep-th/0205052}{ hep-th/0205052}.

\bibitem{Hartnoll:2008vx}
S.~A. Hartnoll, C.~P. Herzog, and G.~T. Horowitz, ``{Building a Holographic
  Superconductor},'' {\em Phys. Rev. Lett.} {\bf 101} (2008) 031601,
  \href{http://xxx.lanl.gov/abs/0803.3295}{ 0803.3295}.

\bibitem{Herzog:2008he}
C.~P. Herzog, P.~K. Kovtun, and D.~T. Son, ``{Holographic model of
  superfluidity},'' \href{http://xxx.lanl.gov/abs/0809.4870}{ 0809.4870}.

\bibitem{Fujita:2009kw}
M.~Fujita, W.~Li, S.~Ryu, and T.~Takayanagi, ``{Fractional Quantum Hall Effect
  via Holography: Chern- Simons, Edge States, and Hierarchy},''
  \href{http://xxx.lanl.gov/abs/0901.0924}{ 0901.0924}.

\bibitem{Contino:2004vy}
R.~Contino and A.~Pomarol, ``{Holography for fermions},'' {\em JHEP} {\bf 11}
  (2004) 058, \href{http://xxx.lanl.gov/abs/hep-th/0406257}{ hep-th/0406257}.

\bibitem{Burgess:1993mg}
C.~P. Burgess, S.~Godfrey, H.~Konig, D.~London, and I.~Maksymyk, ``{A Global
  fit to extended oblique parameters},'' {\em Phys. Lett.} {\bf B326} (1994)
  276--281, \href{http://xxx.lanl.gov/abs/hep-ph/9307337}{ hep-ph/9307337}.

\bibitem{Burgess:1993vc}
C.~P. Burgess, S.~Godfrey, H.~Konig, D.~London, and I.~Maksymyk, ``{Model
  independent global constraints on new physics},'' {\em Phys. Rev.} {\bf D49}
  (1994) 6115--6147, \href{http://xxx.lanl.gov/abs/hep-ph/9312291}{
  hep-ph/9312291}.

\bibitem{Cacciapaglia:2004jz}
G.~Cacciapaglia, C.~Csaki, C.~Grojean, and J.~Terning, ``{Oblique corrections
  from Higgsless models in warped space},'' {\em Phys. Rev.} {\bf D70} (2004)
  075014, \href{http://xxx.lanl.gov/abs/hep-ph/0401160}{ hep-ph/0401160}.

\bibitem{Amsler:2008zz}
{\bf Particle Data Group} Collaboration, C.~Amsler {\em et.~al.}, ``{Review of
  particle physics},'' {\em Phys. Lett.} {\bf B667} (2008) 1.

\bibitem{Csaki:2002gy}
C.~Csaki, J.~Erlich, and J.~Terning, ``{The effective Lagrangian in the
  Randall-Sundrum model and electroweak physics},'' {\em Phys. Rev.} {\bf D66}
  (2002) 064021, \href{http://xxx.lanl.gov/abs/hep-ph/0203034}{
  hep-ph/0203034}.

\bibitem{Agashe:2007mc}
K.~Agashe, C.~Csaki, C.~Grojean, and M.~Reece, ``{The S-parameter in
  holographic technicolor models},'' {\em JHEP} {\bf 12} (2007) 003,
  \href{http://xxx.lanl.gov/abs/0704.1821}{ 0704.1821}.

\bibitem{Hirn:2007we}
J.~Hirn, A.~Martin, and V.~Sanz, ``{Benchmarks for new strong interactions at
  the LHC},'' {\em JHEP} {\bf 05} (2008) 084,
  \href{http://xxx.lanl.gov/abs/0712.3783}{ 0712.3783}.

\bibitem{Hirn:2008tc}
J.~Hirn, A.~Martin, and V.~Sanz, ``{Describing viable technicolor scenarios},''
  {\em Phys. Rev.} {\bf D78} (2008) 075026,
  \href{http://xxx.lanl.gov/abs/0807.2465}{ 0807.2465}.

\bibitem{Chivukula:2005bn}
R.~S. Chivukula, E.~H. Simmons, H.-J. He, M.~Kurachi, and M.~Tanabashi,
  ``Deconstructed Higgsless models with one-site delocalization,'' {\em Phys.
  Rev.} {\bf D71} (2005) 115001, \href{http://xxx.lanl.gov/abs/hep-ph/0502162}{
  hep-ph/0502162}.

\bibitem{Chivukula:2005xm}
R.~S. Chivukula, E.~H. Simmons, H.-J. He, M.~Kurachi, and M.~Tanabashi,
  ``{Ideal fermion delocalization in Higgsless models},'' {\em Phys. Rev.} {\bf
  D72} (2005) 015008, \href{http://xxx.lanl.gov/abs/hep-ph/0504114}{
  hep-ph/0504114}.

\bibitem{Chivukula:2006cg}
R.~S. Chivukula {\em et.~al.}, ``{A three site higgsless model},'' {\em Phys.
  Rev.} {\bf D74} (2006) 075011, \href{http://xxx.lanl.gov/abs/hep-ph/0607124}{
  hep-ph/0607124}.

\bibitem{CDFnote}
{CDF Collaboration}, ``Searching for Anomalous Production of $Z$-Bosons with
  High Transverse Momentum in 0.94 $fb^{-1}$ at the Tevatron,''. CDF Note 8452.

\bibitem{hep-ph/0106251}
K.-m. Cheung, ``Constraints on electron quark contact interactions and
  implications to models of leptoquarks and extra Z bosons,'' {\em Phys. Lett.}
  {\bf B517} (2001) 167--176, \href{http://xxx.lanl.gov/abs/hep-ph/0106251}{
  hep-ph/0106251}.

\bibitem{:2007sb}
{\bf CDF} Collaboration, T.~Aaltonen {\em et.~al.}, ``Search for new physics in
  high mass electron-positron events in $p \bar{p}$ collisions at $\sqrt{s}$ =
  1.96-TeV,'' {\em Phys. Rev. Lett.} {\bf 99} (2007) 171802,
  \href{http://xxx.lanl.gov/abs/arXiv:0707.2524 [hep-ex]}{ arXiv:0707.2524
  [hep-ex]}.

\bibitem{arXiv:0707.2524}
{\bf CDF} Collaboration, T.~Aaltonen {\em et.~al.}, ``Search for new physics in
  high mass electron-positron events in $p \bar{p}$ collisions at $\sqrt{s}$ =
  1.96-TeV,'' {\em Phys. Rev. Lett.} {\bf 99} (2007) 171802,
  \href{http://xxx.lanl.gov/abs/arXiv:0707.2524 [hep-ex]}{ arXiv:0707.2524
  [hep-ex]}.

\bibitem{hep-ex/0611022}
{\bf CDF} Collaboration, A.~Abulencia {\em et.~al.}, ``Search for W' boson
  decaying to electron-neutrino pairs in p anti-p collisions at s**(1/2) =
  1.96-TeV,'' {\em Phys. Rev.} {\bf D75} (2007) 091101,
  \href{http://xxx.lanl.gov/abs/hep-ex/0611022}{ hep-ex/0611022}.

\bibitem{CDFnote2}
{CDF Collaboration}, ``Search for diboson ($W^+W^-$ or $W^{\pm}Z^0$) resonances
  in electron, missing $E_T$ and two jets final state.,''. CDF Note 9730.

\bibitem{Aaltonen:2009qu}
{\bf CDF} Collaboration, T.~Aaltonen {\em et.~al.}, ``{Search for the
  Production of Narrow tb Resonances in 1.9 fb-1 of ppbar Collisions at sqrt(s)
  = 1.96 TeV},'' \href{http://xxx.lanl.gov/abs/0902.3276}{ 0902.3276}.

\bibitem{Agashe:2006at}
K.~Agashe, R.~Contino, L.~Da~Rold, and A.~Pomarol, ``{A custodial symmetry for
  Z b anti-b},'' {\em Phys. Lett.} {\bf B641} (2006) 62--66,
  \href{http://xxx.lanl.gov/abs/hep-ph/0605341}{ hep-ph/0605341}.

\bibitem{Katz:2005au}
E.~Katz, A.~E. Nelson, and D.~G.~E. Walker, ``{The intermediate Higgs},'' {\em
  JHEP} {\bf 08} (2005) 074, \href{http://xxx.lanl.gov/abs/hep-ph/0504252}{
  hep-ph/0504252}.

\bibitem{Agashe:2005dk}
K.~Agashe and R.~Contino, ``{The minimal composite Higgs model and electroweak
  precision tests},'' {\em Nucl. Phys.} {\bf B742} (2006) 59--85,
  \href{http://xxx.lanl.gov/abs/hep-ph/0510164}{ hep-ph/0510164}.

\bibitem{Contino:2006qr}
R.~Contino, L.~Da~Rold, and A.~Pomarol, ``{Light custodians in natural
  composite Higgs models},'' {\em Phys. Rev.} {\bf D75} (2007) 055014,
  \href{http://xxx.lanl.gov/abs/hep-ph/0612048}{ hep-ph/0612048}.

\bibitem{Cacciapaglia:2006gp}
G.~Cacciapaglia, C.~Csaki, G.~Marandella, and J.~Terning, ``{A New Custodian
  for a Realistic Higgsless Model},'' {\em Phys. Rev.} {\bf D75} (2007) 015003,
  \href{http://xxx.lanl.gov/abs/hep-ph/0607146}{ hep-ph/0607146}.

\bibitem{Contino:2008hi}
R.~Contino and G.~Servant, ``{Discovering the top partners at the LHC using
  same-sign dilepton final states},'' {\em JHEP} {\bf 06} (2008) 026,
  \href{http://xxx.lanl.gov/abs/0801.1679}{ 0801.1679}.

\bibitem{Brooijmans:2008se}
G.~Brooijmans {\em et.~al.}, ``{New Physics at the LHC: A Les Houches Report.
  Physics at Tev Colliders 2007 -- New Physics Working Group},''
  \href{http://xxx.lanl.gov/abs/0802.3715}{ 0802.3715}.

\bibitem{Barbieri:2003pr}
R.~Barbieri, A.~Pomarol, and R.~Rattazzi, ``{Weakly coupled Higgsless theories
  and precision electroweak tests},'' {\em Phys. Lett.} {\bf B591} (2004)
  141--149, \href{http://xxx.lanl.gov/abs/hep-ph/0310285}{ hep-ph/0310285}.

\bibitem{Lai:1999wy}
{\bf CTEQ} Collaboration, H.~L. Lai {\em et.~al.}, ``{Global QCD analysis of
  parton structure of the nucleon: CTEQ5 parton distributions},'' {\em Eur.
  Phys. J.} {\bf C12} (2000) 375--392,
  \href{http://xxx.lanl.gov/abs/hep-ph/9903282}{ hep-ph/9903282}.

\bibitem{Alwall:2007st}
J.~Alwall {\em et.~al.}, ``MadGraph/MadEvent v4: The New Web Generation,'' {\em
  JHEP} {\bf 09} (2007) 028, \href{http://xxx.lanl.gov/abs/arXiv:0706.2334
  [hep-ph]}{ arXiv:0706.2334 [hep-ph]}.

\bibitem{hep-ph/0703031}
P.~Meade and M.~Reece, ``BRIDGE: Branching ratio inquiry / decay generated
  events,'' \href{http://xxx.lanl.gov/abs/hep-ph/0703031}{ hep-ph/0703031}.

\bibitem{Sjostrand:2006za}
T.~Sjostrand, S.~Mrenna, and P.~Skands, ``PYTHIA 6.4 physics and manual,'' {\em
  JHEP} {\bf 05} (2006) 026, \href{http://xxx.lanl.gov/abs/hep-ph/0603175}{
  hep-ph/0603175}.

\bibitem{PGS}
{Conway et al.}, ``Pretty Good Simulator v 4.0.'' {PGS}.

\bibitem{DeSimone:2009ws}
A.~De~Simone, J.~Fan, V.~Sanz, and W.~Skiba, ``{Leptogenic Supersymmetry},''
  \href{http://xxx.lanl.gov/abs/0903.5305}{ 0903.5305}.

\bibitem{unknown:1999fr}
{ATLAS TDR}, ``ATLAS detector and physics performance. Technical design report.
  Vol. 2,''. CERN-LHCC-99-15.

\bibitem{unknown:2006fr}
{CMS TDR}, ``CMS TDR Vol. 2,''. report CERN/LHCC/2006-001 (2006).

\bibitem{Cvetic:1997wu}
M.~Cvetic and P.~Langacker, ``{Z' physics and supersymmetry},''
  \href{http://xxx.lanl.gov/abs/hep-ph/9707451}{ hep-ph/9707451}.

\bibitem{Langacker:2000ju}
P.~Langacker and M.~Plumacher, ``{Flavor changing effects in theories with a
  heavy $Z^\prime$ boson with family nonuniversal couplings},'' {\em Phys.
  Rev.} {\bf D62} (2000) 013006, \href{http://xxx.lanl.gov/abs/hep-ph/0001204}{
  hep-ph/0001204}.

\bibitem{Chivukula:2002ry}
R.~S. Chivukula and E.~H. Simmons, ``{Electroweak limits on non-universal Z'
  bosons},'' {\em Phys. Rev.} {\bf D66} (2002) 015006,
  \href{http://xxx.lanl.gov/abs/hep-ph/0205064}{ hep-ph/0205064}.

\bibitem{Rizzo:2006nw}
T.~G. Rizzo, ``{$Z^\prime$ phenomenology and the LHC},''
  \href{http://xxx.lanl.gov/abs/hep-ph/0610104}{ hep-ph/0610104}.

\bibitem{Feldman:2006wb}
D.~Feldman, Z.~Liu, and P.~Nath, ``{The Stueckelberg $Z$ Prime at the LHC:
  Discovery Potential, Signature Spaces and Model Discrimination},'' {\em JHEP}
  {\bf 11} (2006) 007, \href{http://xxx.lanl.gov/abs/hep-ph/0606294}{
  hep-ph/0606294}.

\bibitem{Cata:2009iy}
O.~Cata, G.~Isidori, and J.~F. Kamenik, ``{Drell-Yan production of Heavy
  Vectors in Higgsless models},'' \href{http://xxx.lanl.gov/abs/0905.0490}{
  0905.0490}.

\bibitem{hep-ph/0412278}
A.~Birkedal, K.~Matchev, and M.~Perelstein, ``Collider phenomenology of the
  Higgsless models,'' {\em Phys. Rev. Lett.} {\bf 94} (2005) 191803,
  \href{http://xxx.lanl.gov/abs/hep-ph/0412278}{ hep-ph/0412278}.

\bibitem{He:2007ge}
H.-J. He {\em et.~al.}, ``{LHC Signatures of New Gauge Bosons in Minimal
  Higgsless Model},'' {\em Phys. Rev.} {\bf D78} (2008) 031701,
  \href{http://xxx.lanl.gov/abs/0708.2588}{ 0708.2588}.

\bibitem{Accomando:2008dm}
E.~Accomando, S.~De~Curtis, D.~Dominici, and L.~Fedeli, ``{The four site
  Higgsless model at the LHC},'' {\em Nuovo Cim.} {\bf 123B} (2008) 809--811,
  \href{http://xxx.lanl.gov/abs/0807.2951}{ 0807.2951}.

\bibitem{Accomando:2008jh}
E.~Accomando, S.~De~Curtis, D.~Dominici, and L.~Fedeli, ``{Drell-Yan production
  at the LHC in a four site Higgsless model},''
  \href{http://xxx.lanl.gov/abs/0807.5051}{ 0807.5051}.

\bibitem{Belyaev:2008yj}
A.~Belyaev {\em et.~al.}, ``{Technicolor Walks at the LHC},''
  \href{http://xxx.lanl.gov/abs/0809.0793}{ 0809.0793}.

\bibitem{Piai:2009da}
M.~Piai and M.~Round, ``{Mass-degenerate Heavy Vector Mesons at Hadron
  Colliders},'' \href{http://xxx.lanl.gov/abs/0904.1524}{ 0904.1524}.

\bibitem{Cacciapaglia:2009ic}
G.~Cacciapaglia, A.~Deandrea, and S.~De~Curtis, ``{Nearby resonances beyond the
  Breit-Wigner approximation},'' \href{http://xxx.lanl.gov/abs/0906.3417}{
  0906.3417}.

\bibitem{Heinemeyer:1998yj}
S.~Heinemeyer, W.~Hollik, and G.~Weiglein, ``{FeynHiggs: a program for the
  calculation of the masses of the neutral CP-even Higgs bosons in the MSSM},''
  {\em Comput. Phys. Commun.} {\bf 124} (2000) 76--89,
  \href{http://xxx.lanl.gov/abs/hep-ph/9812320}{ hep-ph/9812320}.

\bibitem{Hirn:2007bb}
J.~Hirn and V.~Sanz, ``{(Not) summing over Kaluza-Kleins},'' {\em Phys. Rev.}
  {\bf D76} (2007) 044022, \href{http://xxx.lanl.gov/abs/hep-ph/0702005}{
  hep-ph/0702005}.

\bibitem{Ohl:2008ri}
T.~Ohl and C.~Speckner, ``{Production of Almost Fermiophobic Gauge Bosons in
  the Minimal Higgsless Model at the LHC},'' {\em Phys. Rev.} {\bf D78} (2008)
  095008, \href{http://xxx.lanl.gov/abs/0809.0023}{ 0809.0023}.

\bibitem{Alves:2009aa}
A.~Alves, O.~J.~P. Eboli, D.~Goncalves, M.~C. Gonzalez-Garcia, and J.~K.
  Mizukoshi, ``{Signals for New Spin-1 Resonances in Electroweak Gauge Boson
  Pair Production at the LHC},'' \href{http://xxx.lanl.gov/abs/0907.2915}{
  0907.2915}.

\bibitem{Eichten:1996dx}
E.~Eichten and K.~D. Lane, ``{Low-scale technicolor at the Tevatron},'' {\em
  Phys. Lett.} {\bf B388} (1996) 803--807,
  \href{http://xxx.lanl.gov/abs/hep-ph/9607213}{ hep-ph/9607213}.

\bibitem{Eichten:1997yq}
E.~Eichten, K.~D. Lane, and J.~Womersley, ``{Finding low-scale technicolor at
  hadron colliders},'' {\em Phys. Lett.} {\bf B405} (1997) 305--311,
  \href{http://xxx.lanl.gov/abs/hep-ph/9704455}{ hep-ph/9704455}.

\bibitem{Alves:2008up}
A.~Alves, O.~J.~P. Eboli, M.~C. Gonzalez-Garcia, and J.~K. Mizukoshi,
  ``{Deciphering the spin of new resonances in Higgsless models},''
  \href{http://xxx.lanl.gov/abs/0810.1952}{ 0810.1952}.

\bibitem{Barbieri:2004qk}
R.~Barbieri, A.~Pomarol, R.~Rattazzi, and A.~Strumia, ``{Electroweak symmetry
  breaking after LEP1 and LEP2},'' {\em Nucl. Phys.} {\bf B703} (2004)
  127--146, \href{http://xxx.lanl.gov/abs/hep-ph/0405040}{ hep-ph/0405040}.

\bibitem{Mangano:2002ea}
M.~L. Mangano, M.~Moretti, F.~Piccinini, R.~Pittau, and A.~D. Polosa, ``ALPGEN,
  a generator for hard multiparton processes in hadronic collisions,'' {\em
  JHEP} {\bf 07} (2003) 001, \href{http://xxx.lanl.gov/abs/hep-ph/0206293}{
  hep-ph/0206293}.

\end{thebibliography}\endgroup
\end{document}